\documentclass{pasj00}

\begin{document}

\SetRunningHead{Y. Shirasaki et al.}
               {Multi-Component Analysis of Time Resolved Spectra of GRB041006}

\Received{2008/02/12}
\Accepted{2008/04/14}

\title{Multiple Component Analysis of Time Resolved Spectra of GRB041006:
A Clue to the Nature of Underlying Soft Component of GRBs.}

%
%

\author{
Yuji \textsc{Shirasaki}, \altaffilmark{1}
Atsumasa \textsc{Yoshida}, \altaffilmark{2,4}
Nobuyuki \textsc{Kawai}, \altaffilmark{3,4}
Toru \textsc{Tamagawa}, \altaffilmark{4}
Takanori \textsc{Sakamoto}, \altaffilmark{5}
}

\author{
Motoko \textsc{Suzuki}, \altaffilmark{6}
Yujin \textsc{Nakagawa}, \altaffilmark{2}
Akina \textsc{Kobayashi}, \altaffilmark{2}
Satoshi \textsc{Sugita}, \altaffilmark{2,4}
Ichiro \textsc{Takahashi}, \altaffilmark{2}
}

\author{
Makoto \textsc{Arimoto}, \altaffilmark{3}
Takashi \textsc{Shimokawabe} \altaffilmark{3}
Nicolas Vasquez \textsc{Pazmino}, \altaffilmark{3}
Takuto \textsc{Ishimura}, \altaffilmark{3}
Rie \textsc{Sato}, \altaffilmark{7}
}

\author{
Masaru \textsc{Matsuoka}, \altaffilmark{6}
Edward \textsc{E. Fenimore}, \altaffilmark{8}
Mark \textsc{Galassi}, \altaffilmark{8}
Donald \textsc{Q. Lamb}, \altaffilmark{9}
Carlo \textsc{Graziani}, \altaffilmark{9}
}

\author{
Timothy \textsc{Q. Donaghy}, \altaffilmark{9}
Jean-Luc \textsc{Atteia}, \altaffilmark{10}
Alexandre \textsc{Pelangeon}, \altaffilmark{10}
Roland \textsc{Vanderspek}, \altaffilmark{11}
}

\author{
Geoffrey \textsc{B. Crew}, \altaffilmark{11}
John \textsc{P. Doty}, \altaffilmark{11}
Joel \textsc{Villasenor}, \altaffilmark{11}
Gregory \textsc{Prigozhin}, \altaffilmark{11}
Nat \textsc{Butler}, \altaffilmark{11,12}
}

\author{
George  \textsc{R. Ricker}, \altaffilmark{11}
Kevin \textsc{Hurley}, \altaffilmark{12}
Stanford \textsc{E. Woosley}, \altaffilmark{13}
and
Graziella \textsc{Pizzichini} \altaffilmark{14}
}

\altaffiltext{1}{National Astronomical Observatory of Japan, Osawa,
Mitaka, Tokyo, 181-8588}
\email{yuji.shirasaki@nao.ac.jp}

\altaffiltext{2}{Department of Physics and Mathematics, Aoyama Gakuin
University, \\5-10-1 Fuchinobe, Sagamihara, Kanagawa 229-8558}

\altaffiltext{3}{Department of Physics, Tokyo Institute of Technology,
2-12-1 Ookayama, \\Meguro-ku, Tokyo 152-8551}

\altaffiltext{4}{RIKEN, 2-1 Hirosawa, Wako Saitama 351-0198}

\altaffiltext{5}{Goddard Space Flight Center, NASA, Greenbelt,
Maryland, 20771, USA}

\altaffiltext{6}{JAXA, 2-1-1 Sengen, Tsukuba, Ibaraki, 305-8505}

\altaffiltext{7}{JAXA/ISAS, 3-1-1 Yoshinodai, Sagamihara, Kanagawa, 229-8510}

\altaffiltext{8}{Los Alamos National Laboratory, P. O. Box 1663, Los
Alamos, NM, 87545, USA}

\altaffiltext{9}{Department of Astronomy and Astrophysics, University of
Chicago, \\5640 South Ellis Avenue, Chicago, Illinois 60637, USA}

\altaffiltext{10}{LATT,  Universit{\'e} de Toulouse, CNRS, 14 avenue E. 
Belin, 31400 Toulouse, France}

\altaffiltext{11}{Center for Space Research, MIT, 77 Vassar Street,
Cambridge, Massachusetts, 02139-4307, USA}

\altaffiltext{12}{Space Sciences Laboratory, 7 Gauss Way, University of California ,
Berkeley, California, 94720-7450}

\altaffiltext{13}{Department of Astronomy and Astrophysics, University
of California at Santa Cruz, \\477 Clark Kerr Hall, Santa Cruz, California,
95064, USA}

\altaffiltext{14}{INAF/IASF Bologna, Via Gobetti 101, 40129 Bologna,
Italy}


%

\KeyWords{gamma-rays:busts --- X-rays: bursts ---  X-rays: individual (GRB041006)} 

\maketitle

\begin{abstract}
GRB 041006 was detected by HETE-2 at 12:18:08 UT on 06 October 2004.
This GRB displays a soft X-ray emission, a precursor before the onset of the 
main event, and also a soft X-ray tail after the end of the main peak.
The light curves in four different energy bands display different features;
At higher energy bands several peaks are seen in the light curve,
while at lower energy bands a single broader bump dominates.
It is expected that these different features are the result of a mixture
of several components each of which has different energetics and variability.
To reveal the nature of each component, we analysed the time resolved 
spectra and they are successfully resolved into several components.
We also found that these components can be classified into two distinct
classes; 
One is a component which has an exponential decay of $E_{p}$ with
a characteristic timescale shorter than $\sim$~30~sec, and its spectrum 
is well represented by a broken power law function, which is frequently 
observed in many prompt GRB emissions, so it should have an internal-shock 
origin.
Another is a component whose $E_{p}$ is almost unchanged with characteristic
timescale longer than $\sim$~60~sec, and shows a very soft emission and 
slower variability.
The spectrum of the soft component is characterized by either a broken power
law or a black body spectrum.
This component might originate from a relatively wider and lower velocity 
jet or a photosphere of the fireball.
By assuming that the soft component is a thermal emission, the radiation 
radius is initially  $4.4 \times 10^{6}$~km, which is a 
typical radius of a blue supergiant, and its expansion velocity is
$2.4 \times 10^{5}$~km s$^{-1}$ in the source frame.

\end{abstract}

%

%

\section{Introduction}

On October 6, 2004 the High Energy Transient Explorer~2 (HETE-2) detected a 
gamma-ray burst (GRB) with soft X-ray emission before the onset of the main 
event.
Such soft emission, a precursor, is predicted in some of theoretical models.
The fireball undergoes a transition from an optically thick phase to an 
optically thin phase, and thermal radiation (the fireball precursor) may
occur during this transition (\cite{Paczynsky1986}; \cite{Daigne2002}).
%
%
A precursor (progenitor precursor) may also be emitted by the interaction of 
the jet with the progenitor star (\cite{Ramirez2002}; \cite{Waxman2003}).
The external shock by the first relativistic shell can also produce the
non-thermal precursor (\cite{Umeda2005}).

Soft precursors are occasionally detected in long GRBs.
The first detection was made by the GINGA satellite (GRB 900126; \cite{Murakami1991}).
In more recent observations, 
the BeppoSAX (e.g. GRB~011121; \cite{Piro2005}),
HETE2 (e.g. GRB~030329; \cite{Vanderspek2004}) and 
Swift (e.g GRB~050820A; \cite{Cenko2006}, GRB~060124; \cite{Romano2006}, 
GRB~061121; \cite{Page2007}) satellites have also detected
precursors.
%
\cite{Lazzati2005} studied bright long BATSE GRB light curves and found that 
in 20\% of the cases there is evidence for soft emission before the main
event.

The precursor is usually detected as a single pulse that is well separated 
in time from the main event, typically several seconds to hundreds of seconds.
The precursor of GRB~041006 is not well separated from the main event and is 
likely to be continuously active during the whole prompt GRB phase.
Such a long lasting soft component was also observed in GRB~030329 
(\cite{Vanderspek2004}).
\cite{Vetere2006} found that for some of the GRBs detected
by the BeppoSAX, there is a slowly varying soft component underlying the 
highly variable main event.
\cite{Borgonovo2007} analyzed the light curves obtained by BATSE, 
Konus, and BeppoSAX, and found that the width of the auto-correlation 
function shows a remarkable bimodal distribution in the rest-frame 
of the source.
This result suggests that there exists a slowly varying soft component in some GRBs.
The relation between the underlying soft X-ray component, the X-ray 
precursor, and the main event is still open to question.

In this paper, we present the results of multiple component analysis of the
time resolved spectra of GRB~041006.
Throughout this paper the peak energies are in the observer's frame,
and quoted errors are at 90\% C.L., unless specified otherwise.
%

%
%
%
%
%
%
%

\section{Observation}

GRB~041006 was detected with the HETE FREGATE 
(\cite{Atteia2003}) and the WXM (\cite{Shirasaki2003}) instruments 
at 12:18:08 UT on 06 October 2004 (\cite{Galassi2004}).
The WXM flight software localized the burst in real time, resulting in a 
GCN Notice 42 seconds after the burst trigger. 
The prompt error region was a circle of 14 arcminute radius 
(90\% confidence) centered at 
\timeform{00h54m54s}, \timeform{+01D18'37"}
(J2000).
Ground analyses of the burst data allowed the error region to be refined 
to a circle of 5.0 arcminute radius (90\% confidence) centered at 
\timeform{00h54m53s}, \timeform{+01D12'04"}
(J2000).

1.4 hours after the trigger, the optical afterglow was found by 
\cite{Costa2004}, and the redshift was first reported by \cite{Fugazza2004}
and later confirmed by \cite{Price2004} to be $z = 0.716$.
Follow-up observations were made at various observation sites
(e.g. \cite{Urata2007}).
VLA observations were made but no radio sources were detected 
(\cite{Soderberg2004}).
%
%
%
The X-ray afterglow was found by \cite{Butler2005}, and it exhibited a 
power law decay with a slope of $-1.0 \pm 0.1$.
The X-ray spectrum was characterized by an absorbed power law model with a photon index
of $\Gamma = 1.9 \pm 0.2$ and n$_{\rm H}$ $= (1.1 \pm 0.5) \times 10^{21}$ cm$^{-2}$. 
The emergence of a supernova component was reported  by \cite{Bikmaev2004} 
and \cite{Garg2004}.
The field of GRB 041006 was imaged by \cite{Soderberg2006} using the WFC of the 
ACS on-board HST, and they found a SN 1998bw-like supernova dimmed by
$\sim$0.3 magnitudes.

\section{Analysis}

The data obtained by the WXM and FREGATE instruments were reduced 
and calibrated in the standard manner.
We used WXM TAG data and FREGATE PH data.

\subsection{Temporal Properties}

\begin{figure*}
\begin{center}
    \FigureFile(0.8\textwidth,\textwidth){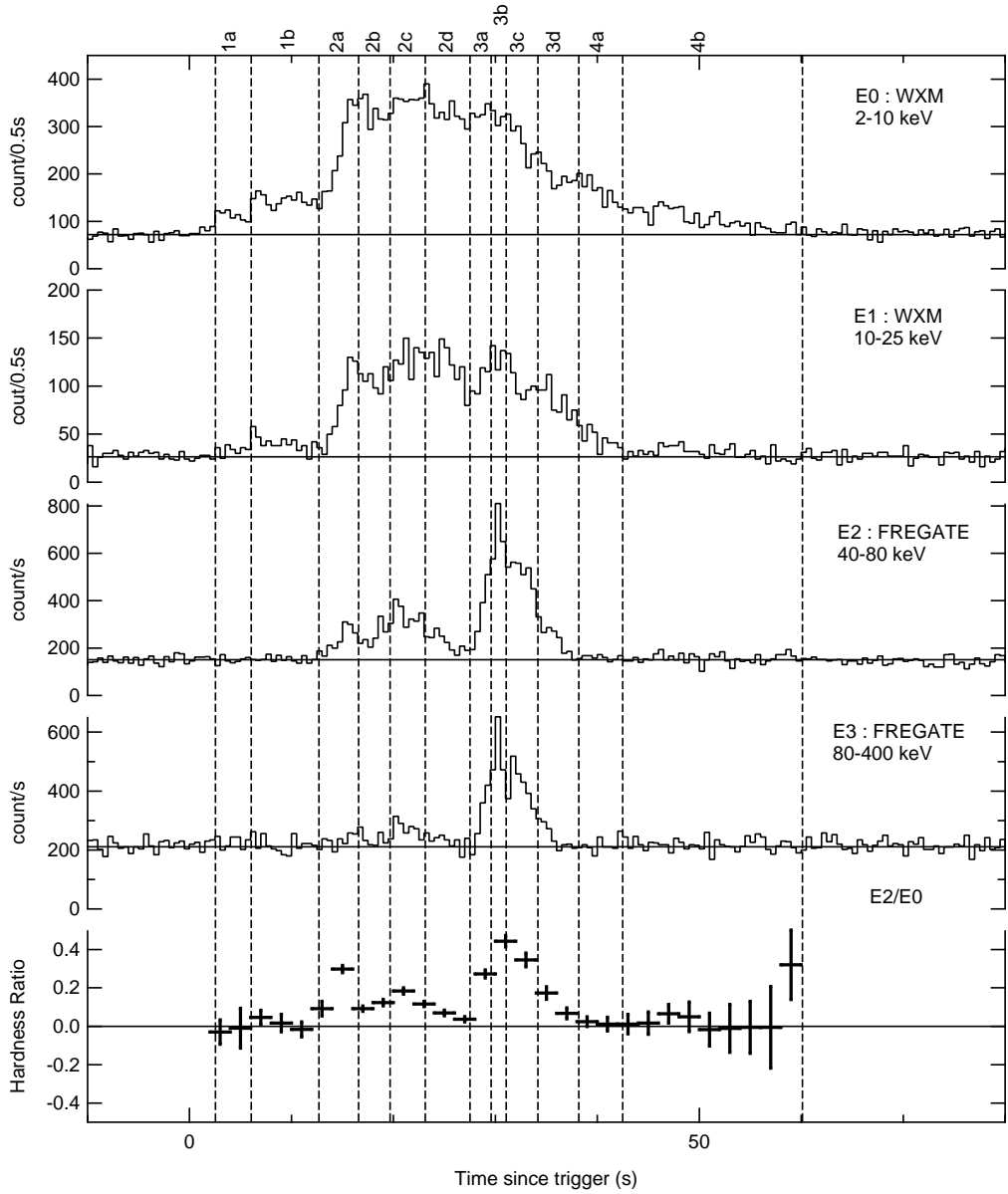}
\caption{Light curves of GRB041006 in four energy bands and hardness ratio.
From top to bottom, 2--10~keV, 10--25~keV, 40--80~keV, and 80--400~keV.
The hardness ratio is calculated by dividing the 40--80~keV count rate
by the 2--10~keV count rate. The vertical lines represent
the boundaries of the time intervals for time resolved spectral analysis.
}
\label{fig:LightCurves}
\end{center}
\end{figure*}

Figure~\ref{fig:LightCurves} shows the light curves of GRB~041006 in
four energy bands with 0.5~sec time resolution.
$T_{50}$ and $T_{90}$ are measured for each energy band, and they are shown 
in table~\ref{tbl:T50and90}.
The burst can be divided into four major intervals
according to spectral features, and each major interval is divided 
into a few sub-intervals for time-resolved spectral analysis.
The time intervals for each sub-interval are shown in table~\ref{tbl:timeRegion}.
In interval~1 soft emission showing no prominent activity above 
40~keV occurs, then harder emissions follow in intervals~2 
and 3. 
In interval~4, the hard emission almost disappears and only
gradually decaying soft emission is present.

We call the emission seen in interval~1 an X-ray precursor.
The precursor shows a structured light curve in the lowest energy band
(2--10 keV), which indicates that two emissions are occurring 
successively .
%
%
In interval~2, two peaks are seen in the higher energy bands 
($>$ 40~keV).
The time history of the hardness ratio also clearly shows the corresponding 
peaks.
In the lowest energy bands ($<$ 10~keV), structured emission is 
not clearly seen.
In interval~3, two harder peaks are seen in the highest energy band
(80--400~keV), and this structure is less distinct in the lower 
energy bands.
The emission in interval~4, which we call an X-ray tail, shows
no prominent structure.

From the dissimilarity of the light curves in the four energy bands,
it is inferred that the total emission is composed of several independent 
emissions which have different characteristic energies.
For an example, two components which contribute to the precursor,
four components seen as a peak in the energy bands 40--80~keV
and 80--400~keV, and one broad soft component which constitutes 
the major part of the light curve in the lowest energy band.
To investigate this hypothesis, we performed time resolved spectral analysis
based on a multiple-component spectrum model. 

\subsection{Average Spectral Properties}

\begin{figure*}
\begin{center}
\begin{tabular}{c}
\includegraphics[width=0.8\textwidth]{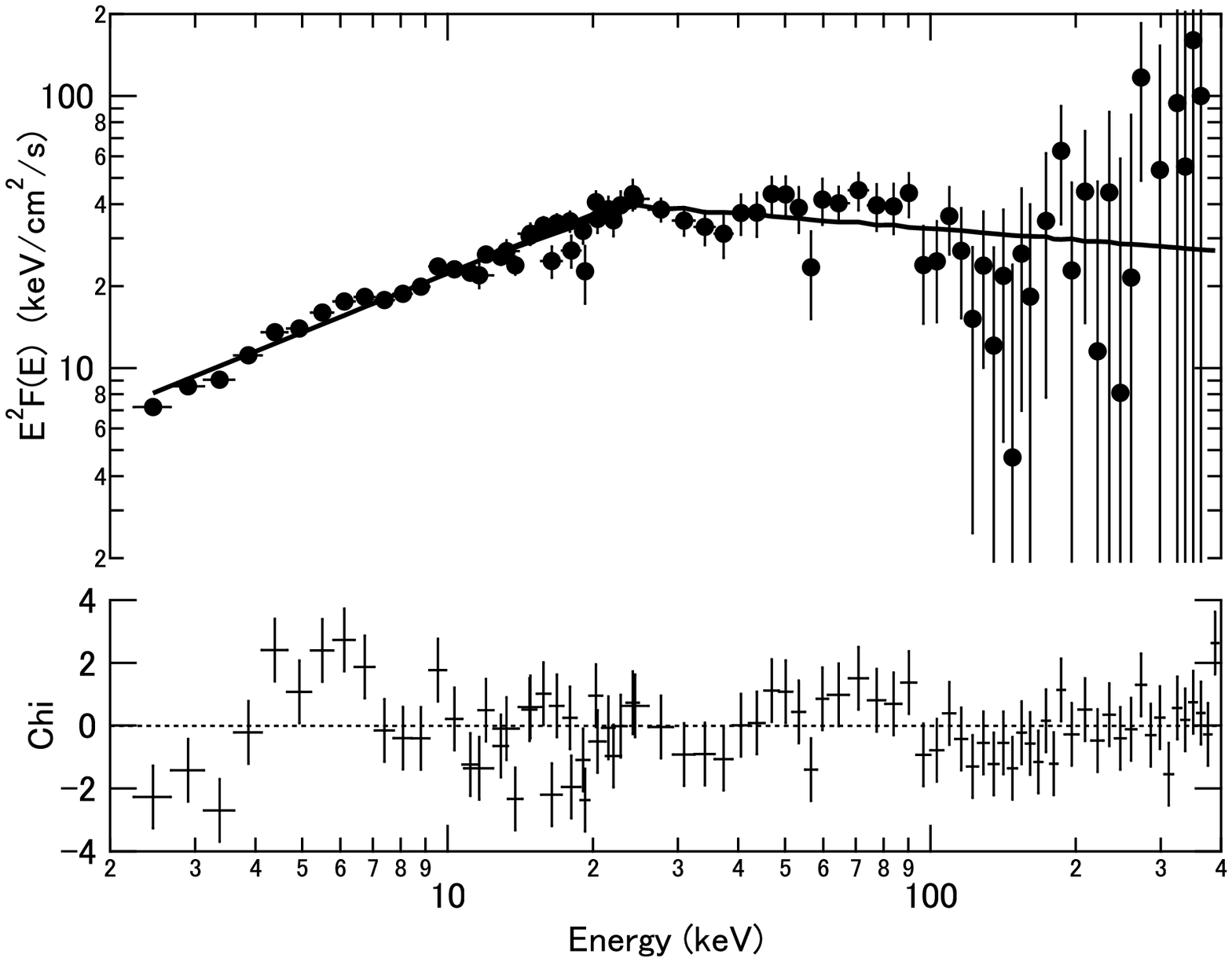}\\
\includegraphics[width=0.8\textwidth]{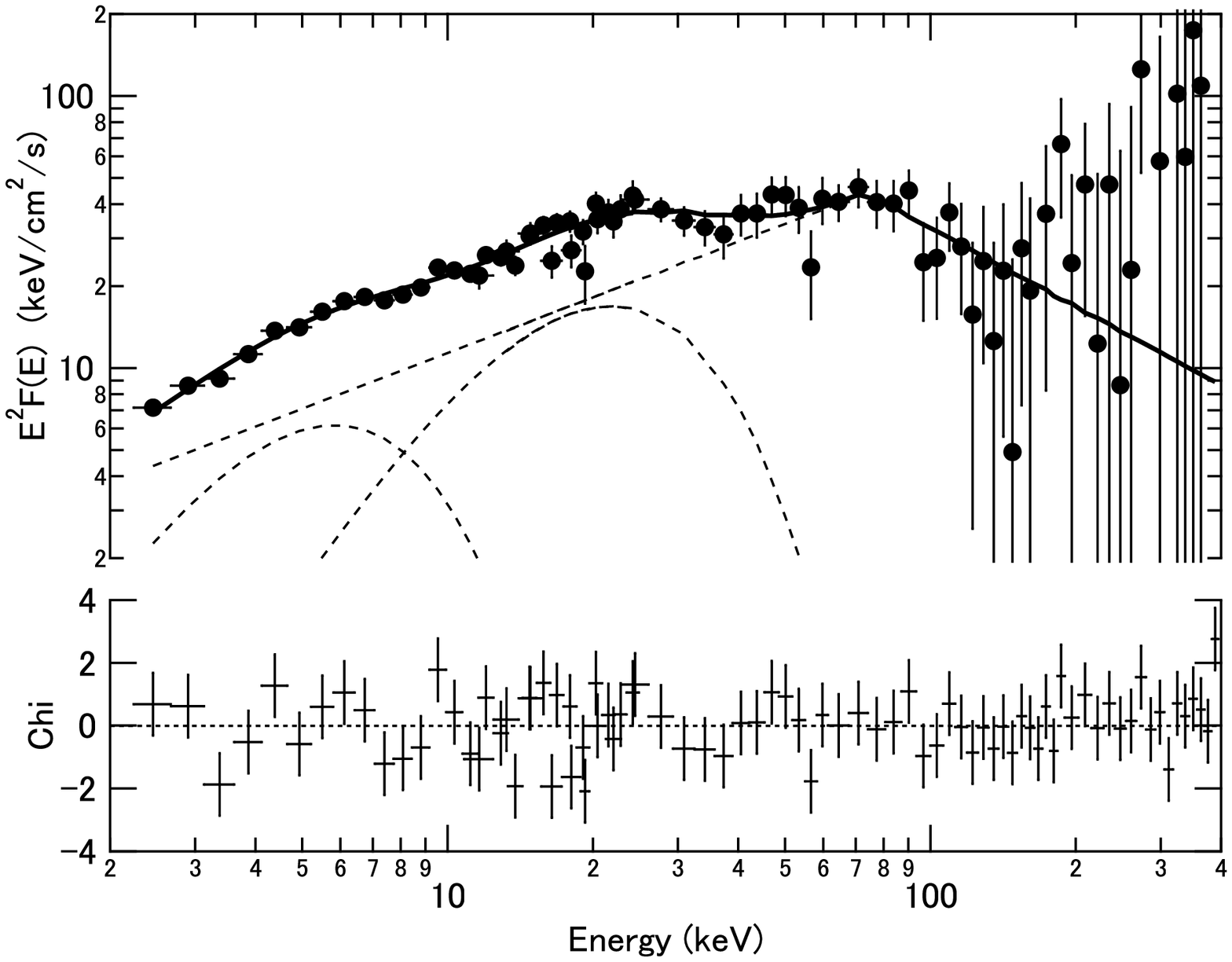}\\
\end{tabular}
\caption{Time averaged unfolded spectrum expressed in 
$\nu f_{\nu}$. 
Top: Fitting result for the broken power law model.
Bottom: Fitting result for the three-component model 
represented by a superposition of one broken power law 
function and two blackbody functions.}
\label{fig:spect_all}
\end{center}
\end{figure*}

The joint spectral analysis of WXM and FREGATE data was 
performed using XSPEC v.11.3.1 (\cite{Arnaud1996}).
The time integrated spectrum of GRB~041006 is approximately
described by a broken power law function (figure~\ref{fig:spect_all});
the low energy photon index is $\alpha = 1.28 \pm 0.02$, 
the high energy index is $\beta = 2.14 \pm 0.07$, 
the break energy is $E_{p} = 22.5 \pm 1.7$~keV and the 
flux at 1~keV is $K = 4.25 \pm 0.15$~cm$^{-2}$s$^{-1}$~keV$^{-1}$,
where the quoted errors are one sigma.
The $\chi^{2}$ is 111.19 for 79 dof, and Null hypothesis 
probability is 0.0099, so the fit is not very good.
From this fitting result, we obtained 
$S_{X} = (5.24 \pm 0.08) \times 10^{-6}$ ergs cm$^{-2}$,
$S_{\gamma} = (7.13 \pm 0.12) \times 10^{-6}$ ergs cm$^{-2}$,
where $S_{X}$ and $S_{\gamma}$ denote fluences in the 
2--30~keV and 30--400~keV energy ranges and the error is 
1 sigma.
As the ratio of fluences is log($S_{x}/S_{\gamma}$) = $-0.13$, 
the GRB can be classified as an X-ray Rich GRB 
(\cite{Sakamoto2005}).

\begin{figure*}
\begin{center}
    \FigureFile(1\textwidth,\textwidth){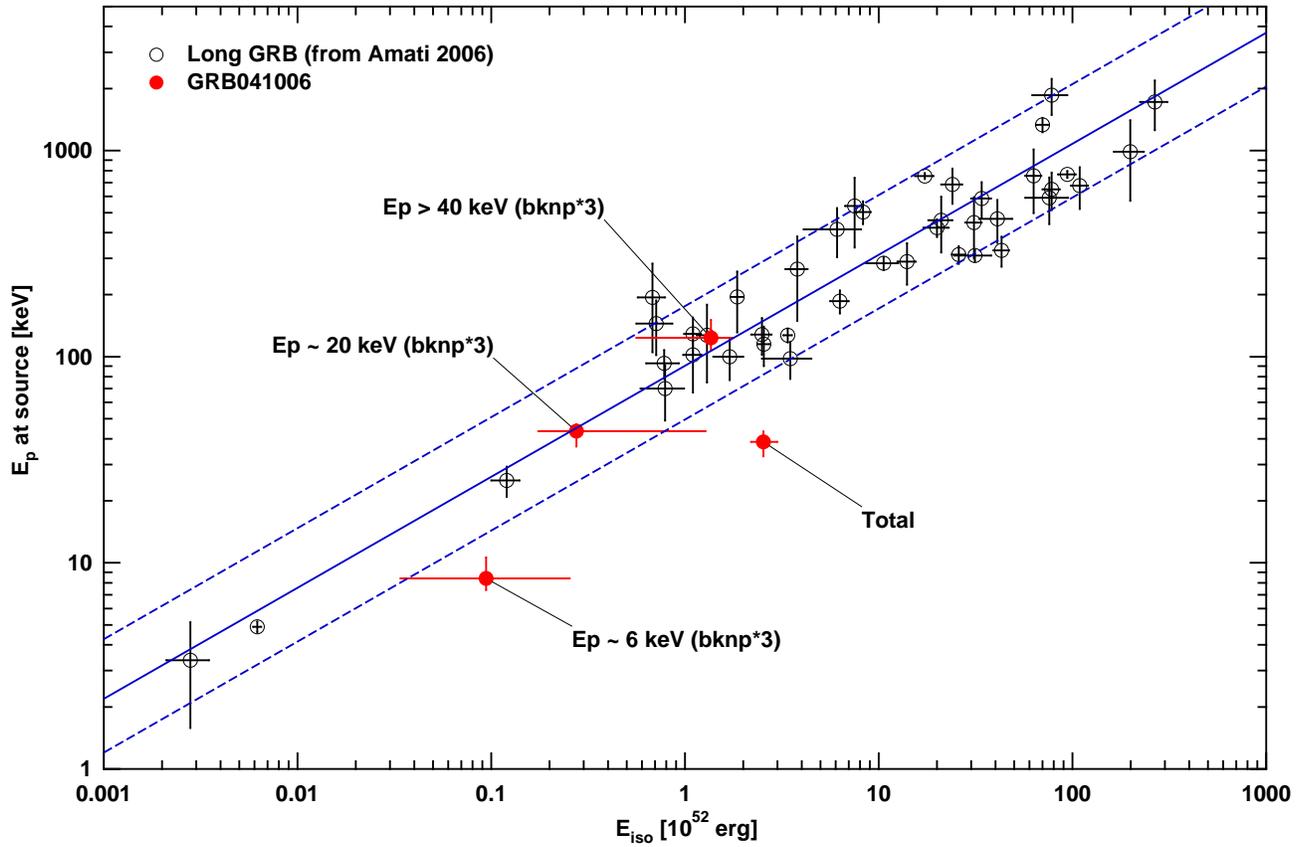}
\caption{$E_{\rm p,src}$-$E_{\rm iso}$ relation for long GRBs. The open circles  
represent the long GRBs compiled by \cite{Amati2006}.
The solid circles represent GRB~041006. 
The solid circle labeled ``Total'' is derived from a single broken 
power law model. 
The other solid circles are derived from a \texttt{bknp*3} model. 
The parameters obtained in this work are summarized in table~\ref{tbl:ep_eiso}.
The solid line represent the average relation
derived from all the points of the open circles, while the dashed lines 
represent lower and upper boundaries which are parallel to the average
relation and contain 90\% of the points.}
\label{fig:amati}
\end{center}
\end{figure*}

The isotropic energy is calculated from:
%
%
\begin{equation}
E_{\rm iso} = \frac{4 \pi D_{L}^{2}}{z + 1} \int_{E_{\rm lo,src}/(z + 1)}^{E_{\rm hi,src}/(z + 1)} E \Phi dE
\end{equation}
where $D_{L}$ is the luminosity distance, $\Phi$ is the
differential photon spectrum, the range of energy intergation is
from 1 keV to 10000 keV in the source frame.
We obtained $E_{\rm iso} = 2.54^{+0.46}_{-0.35} \times 10^{52}$ ergs.
In figure~\ref{fig:amati}, the peak energy in the source 
frame $E_{\rm p, src}$ is plotted against the isotropic energy $E_{\rm iso}$ 
(the point labeled ``Total'').
The relation for GRB~041006 obtained from the one component 
fit is completely outside the Amati relation (\cite{Amati2006}).
Looking at the residual plot in the top panel of figure~\ref{fig:spect_all}, 
an additional soft component is apparently seen around 6~keV
and a systematic excess is also seen around 50--100~keV.
Thus the total spectrum was fitted by a superposition of multiple
basic functions.
As basic functions, we considered a broken power law and a black-body.

For the broken power law model, we used the following function
to estimate the peak energy flux directly:
\begin{eqnarray}
  \label{eq:bknp}
   A(E) & = & K/E_{p}^{2} (E/E_{p})^{-\alpha},\hspace{3em}  E \le E_{p} \\
        &   & K/E_{p}^{2} (E/E_{p})^{-\beta}, \hspace{3em}  E > E_{p} \nonumber
\end{eqnarray}
The parameters $\alpha$ and $\beta$, which are the lower and higher
energy photon indices, are restricted to the range of
-2.0--2.0 and 2.5--5.0, respectively.
The initial value of the break energy $E_{p}$ of the \verb|bknp|
basic function is determined from the local excess of the residual
between the single \verb|bknp| model and the observed data.
The restriction to the break energy $E_{p}$ is applied so 
that the parameter converges around the initial value.

The results of the spectral fit for three three-component 
models are shown in table~\ref{tbl:fit_all}.
For comparison the result of the two-component model and a
fit by the Band function (\cite{Band1993}) and a broken
power law function are also shown in the table.
The fitting parameters for the models \verb|bbody*2+bknp| and
\verb|bknp*3| are given in table~\ref{tbl:fit_params_total}.

Akaike's Information Criterion (AIC) is calculated for each 
model.
AIC (\cite{Akaike1974}) is a very widely used criterion to evaluate 
the goodness of the statistical model from both the goodness of fit 
and the complexity of the model.
AIC is defined by the following equation:
\begin{equation}
   AIC = n \ln{ \left( \frac{\chi^{2}}{n} \right) } + 2 k,
\end{equation}
where $n$ is the number of data points, $k$ is the number of free 
parameters to be estimated, and $\chi^{2}$ is the residual sum of 
squares from the estimated model.
The AIC includes a penalty that is an increasing function of the
number of estimated parameters; overfitting is discouraged, and
thus this method enables one to find the best model for the 
data, with minimum of free parameters.
The model with the lower value of AIC is the one to be preferred.

The most preferable model is \verb|bbody*2+bknp|.
The model name is given by an algebraic expression of the name of
a basic model.
The second most preferable model is \verb|bknp*3|.
The AIC values for the two models are 6.87 and 8.47 respectively.

The lowest AIC does not necessarily select the true model, and the 
degree of the preference is estimated by the AIC difference.
The relation between the degree of the preference and the AIC 
difference ($\Delta_{X}$), however, depends on $n$ and the
models to be compared.
So we evaluate the confidence limit of the AIC difference by 
carrying out a Monte Carlo simulation.
The Monte Carlo simulation was performed by using the \verb|fakeit| 
command of XSPEC, which generated 1000 PHA samples based on the
spectral model to be tested.
For each PHA sample, a spectral fit was performed for both the 
tested model and the model which gave the lowest AIC, and
the AIC difference was calculated.

\begin{figure*}
\begin{center}
\begin{tabular}{cc}
\includegraphics[width=0.45\textwidth]{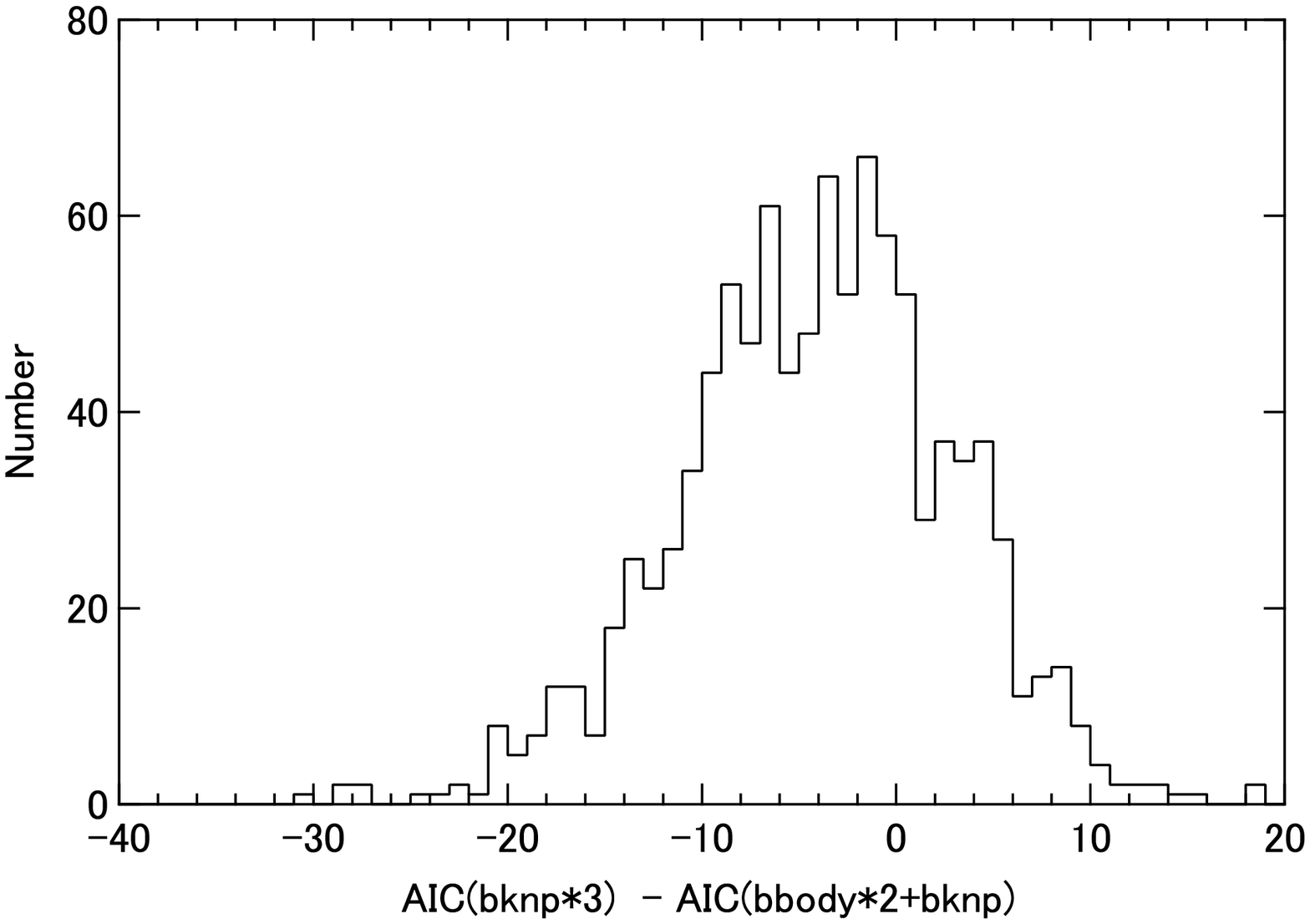}
\includegraphics[width=0.45\textwidth]{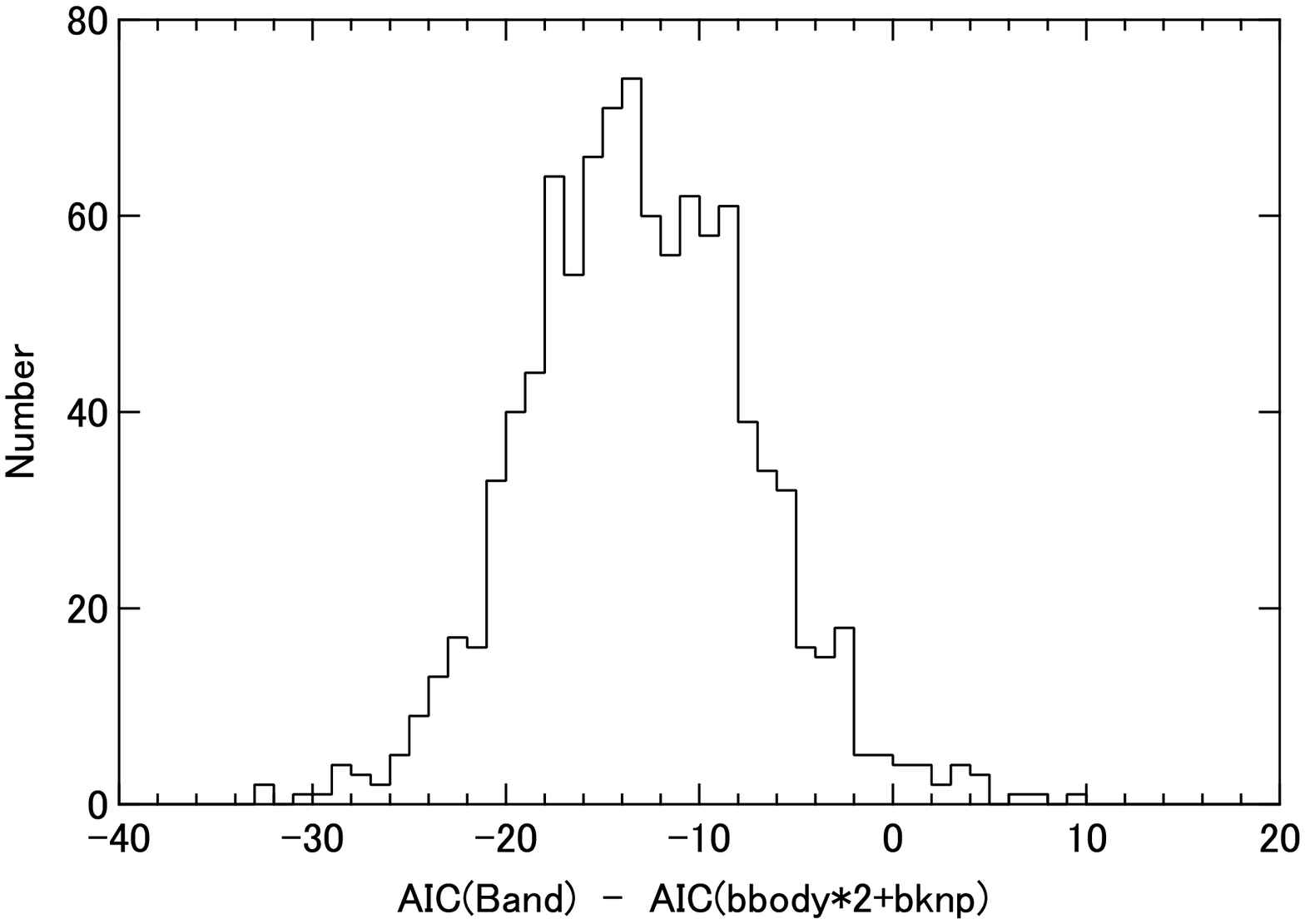}
\end{tabular}
\end{center}
\caption{Left: Simulated distribution of AIC differences between the 
\texttt{bknp*3} and \texttt{bbody*2+bknp} models (left). 
The simulation is performed using the \texttt{bknp*3} model, and 
model fitting to the simulated data is carried out for both the 
models. 
The \texttt{bbody*2+bknp} model is the most preferable model for the
time integrated spectrum.
The AICs for the two models are calculated for each simulated spectrum.
The fraction of events with $\Delta_{AIC} > 0$ corresponds to the 
probability of selecting the wrong model.
Right: Same plot for the Band model.
}
\label{fig:aic_diff}
\end{figure*}

The left panel of figure~\ref{fig:aic_diff} shows a simulated distribution
of the AIC difference 
$\Delta_{\verb|bknp*3|} =$AIC$_{\verb|bknp*3|} - $AIC$_{\verb|bbody*2+bknp|}$.
The simulation was performed with the model spectrum \verb|bknp*3|;
the model parameters were obtained from the fit to the observed
total spectrum.
For each simulated PHA sample, model fit was performed for both the
\verb|bknp*3| model and \verb|bbody*2+bknp|, which is
the most preferred model.
From this result the 90\% confidence limit for $\Delta_{\verb|bknp*3|}$ is 
estimated as 4.7, below which 90\% of samples are included.
The observed AIC difference for the model \verb|bknp*3|
is 2.64, so the model is acceptable at 90\% C.L.
In the case of the Band model (right hand panel of figure~\ref{fig:aic_diff}), 
for 98\% of the samples the AIC is smaller than the most preferred 
model \verb|bbody*2+bknp|.
The observed AIC difference is 13.68, so the Band model is rejected
at higher than 98\% C.L.
All the three three-component models are acceptable at 90\% C.L.
The two-component model is rejected at 90\% C.L.

As the time averaged spectrum of GRB 041006 is well represented by a
superposition of the three components, we examined 
the $E_{\rm p,src}$-$E_{\rm iso}$ relation for each one.
The $E_{\rm iso}$ calculated for a model \verb|bknp*3| are
summarized in table~\ref{tbl:ep_eiso}.
The $E_{\rm iso}$ calculated for a model \verb|bbody*2+bknp| is
also shown in the table for the high energy component.
The result are compared with the other GRBs in figure~\ref{fig:amati}.
The components with $E_{p} > 40$~keV (C) and 
$E_{p} \sim$~20~keV (B) are well within the Amati relation, and
the component $E_{p} \sim $6~keV (A) is out of the 90\% distribution width 
of the Amati relation.
The log($S_{x}/S_{\gamma}$) for the three components are
$-$0.3 for the component C,
0.78 for the component B,
and 0.76 for the component A;
thus they are classified as XRR, XRF and XRF, respectively.

\subsection{Time Resolved Spectral Properties}

Time resolved spectral analysis was performed for 12 independent
time intervals, and also for some intermediate intervals which 
overlap part of one or two adjacent intervals to trace 
the spectral evolution more closely.
We applied multi-component models in the spectral fit, where
the model spectrum is represented as a superposition of an arbitrary
number of basic functions.
The basic functions considered here are black body
(\verb|bbody|), broken power law (\verb|bknp|), 
and a single power law function (\verb|pl|).
The XSPEC built-in model is used for \verb|bbody| and \verb|pl|,
for which the XSPEC model names are \verb|bbodyrad| and \verb|powerlaw|
respectively.
For the broken power law model, we used equation~\ref{eq:bknp}.
%
%
%
%
%
%

\begin{figure*}
\begin{center}
\includegraphics[width=0.7\textwidth]{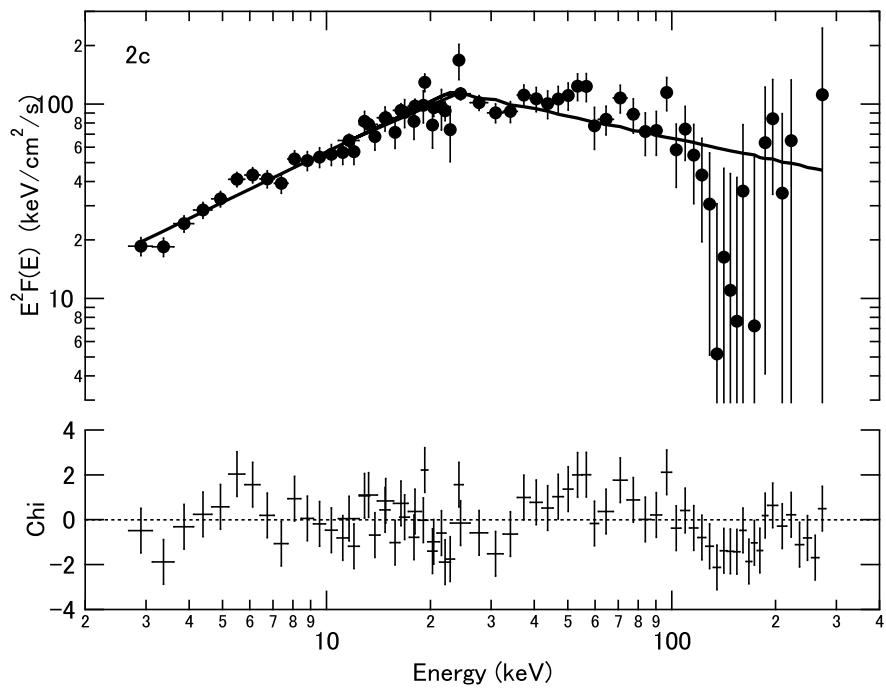}
\end{center}
\caption{An example of spectral fitting for interval~2c, where
a single-component model is used.}
\label{fig:spect_2c}
\end{figure*}

\begin{figure*}
\begin{center}
   \begin{tabular}{cc}
     \FigureFile(0.45\textwidth,0.2\textwidth){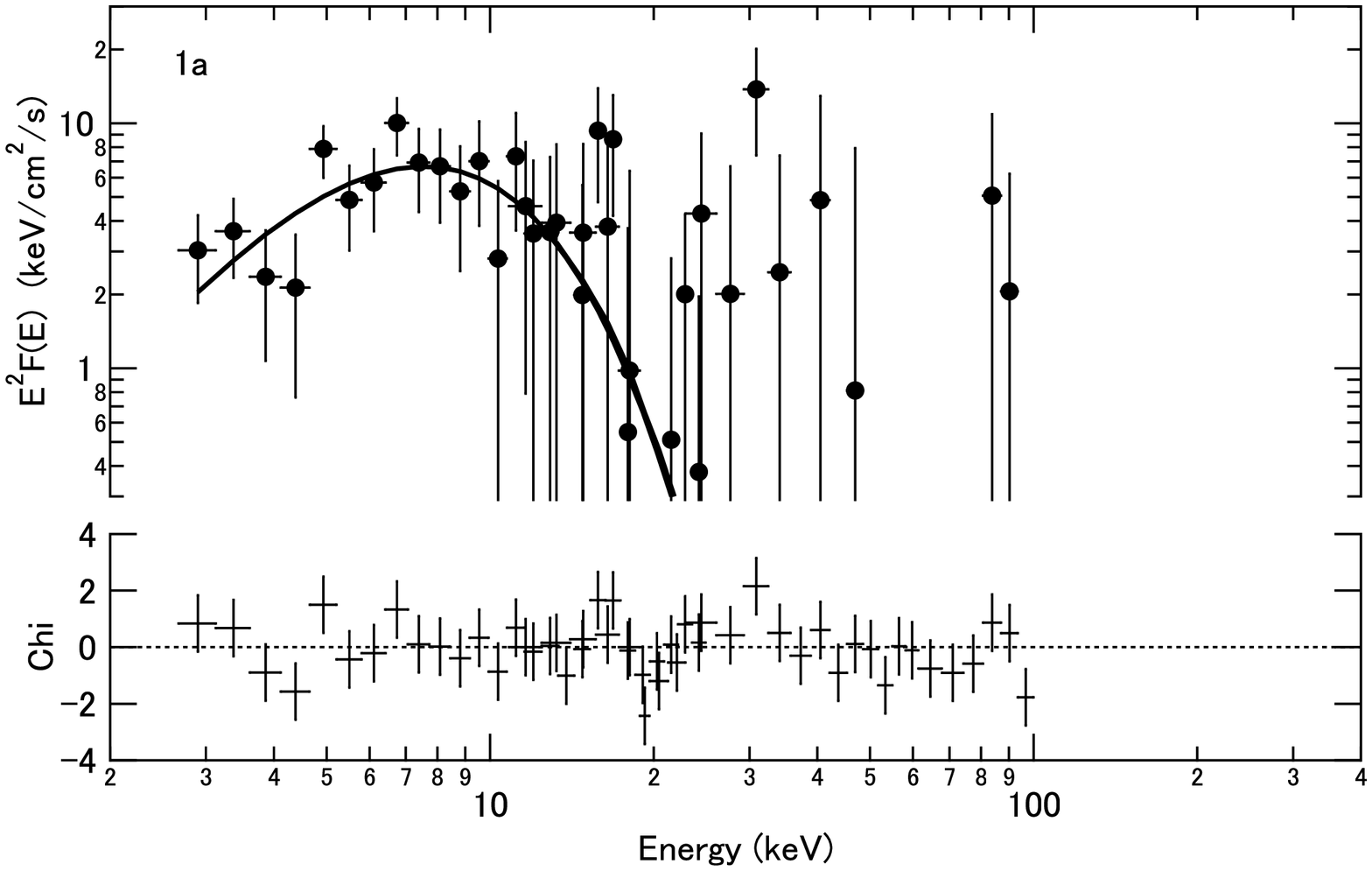} &
     \FigureFile(0.45\textwidth,0.2\textwidth){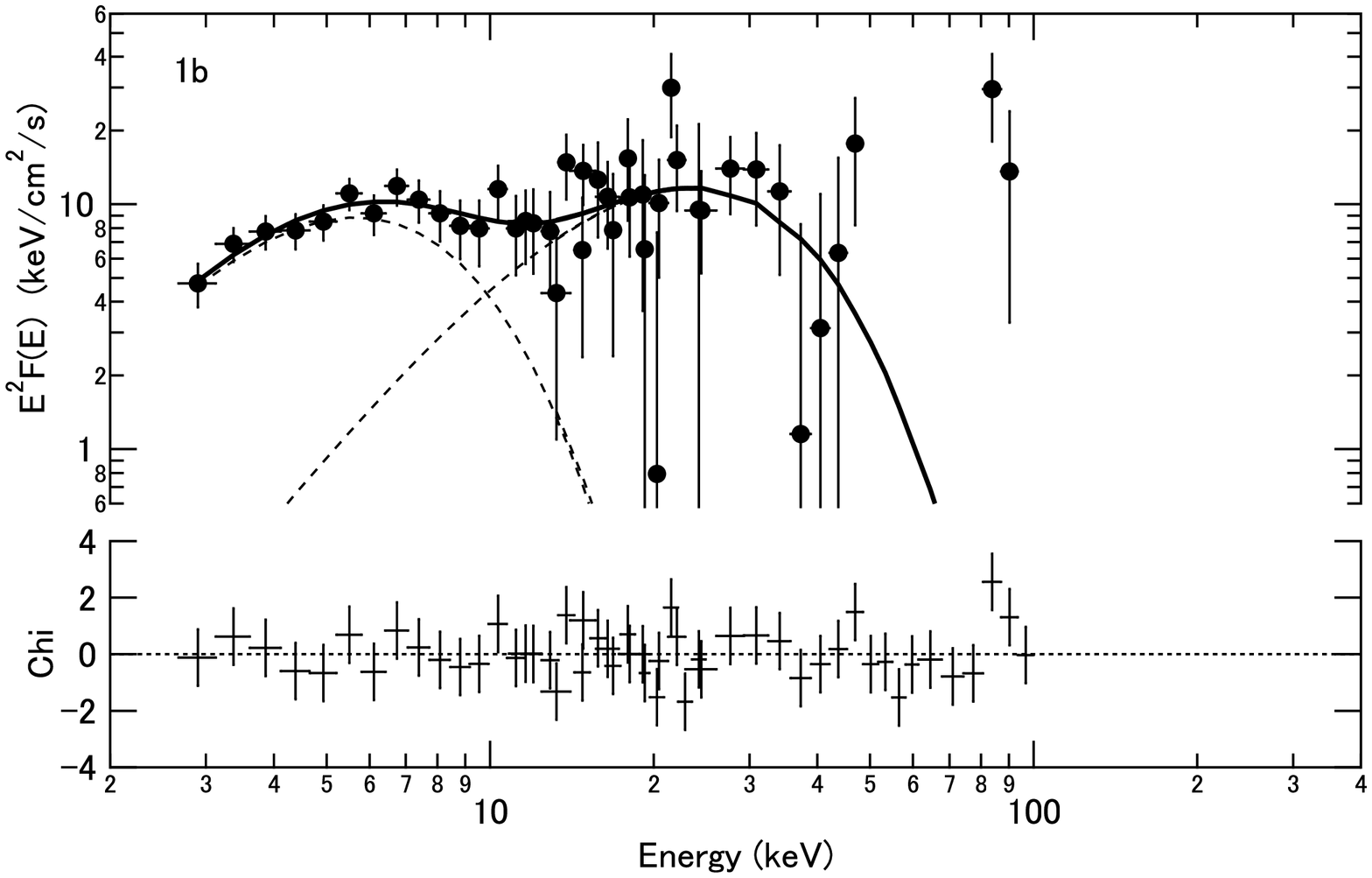} \\
     \FigureFile(0.45\textwidth,0.2\textwidth){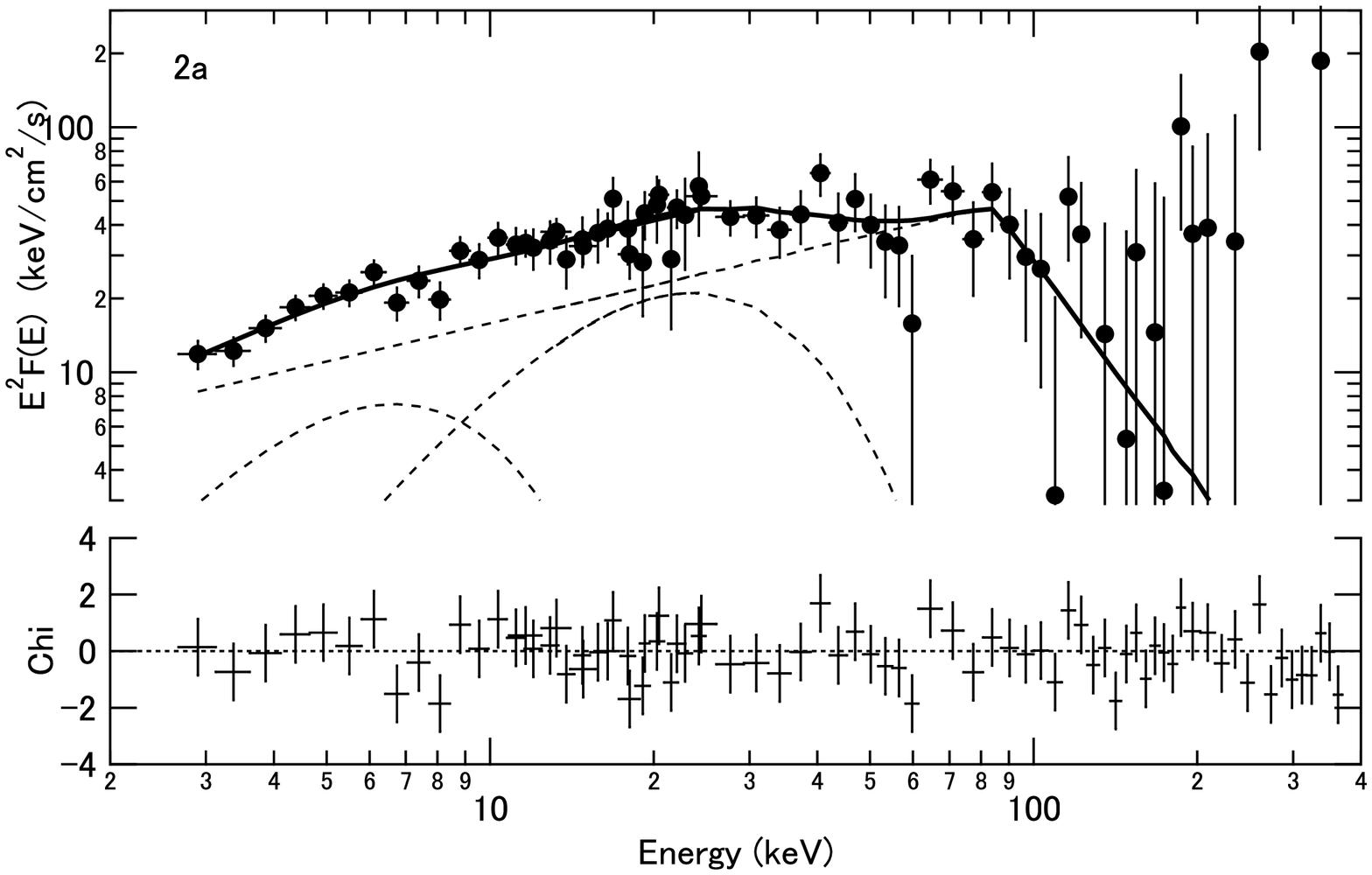} &
     \FigureFile(0.45\textwidth,0.2\textwidth){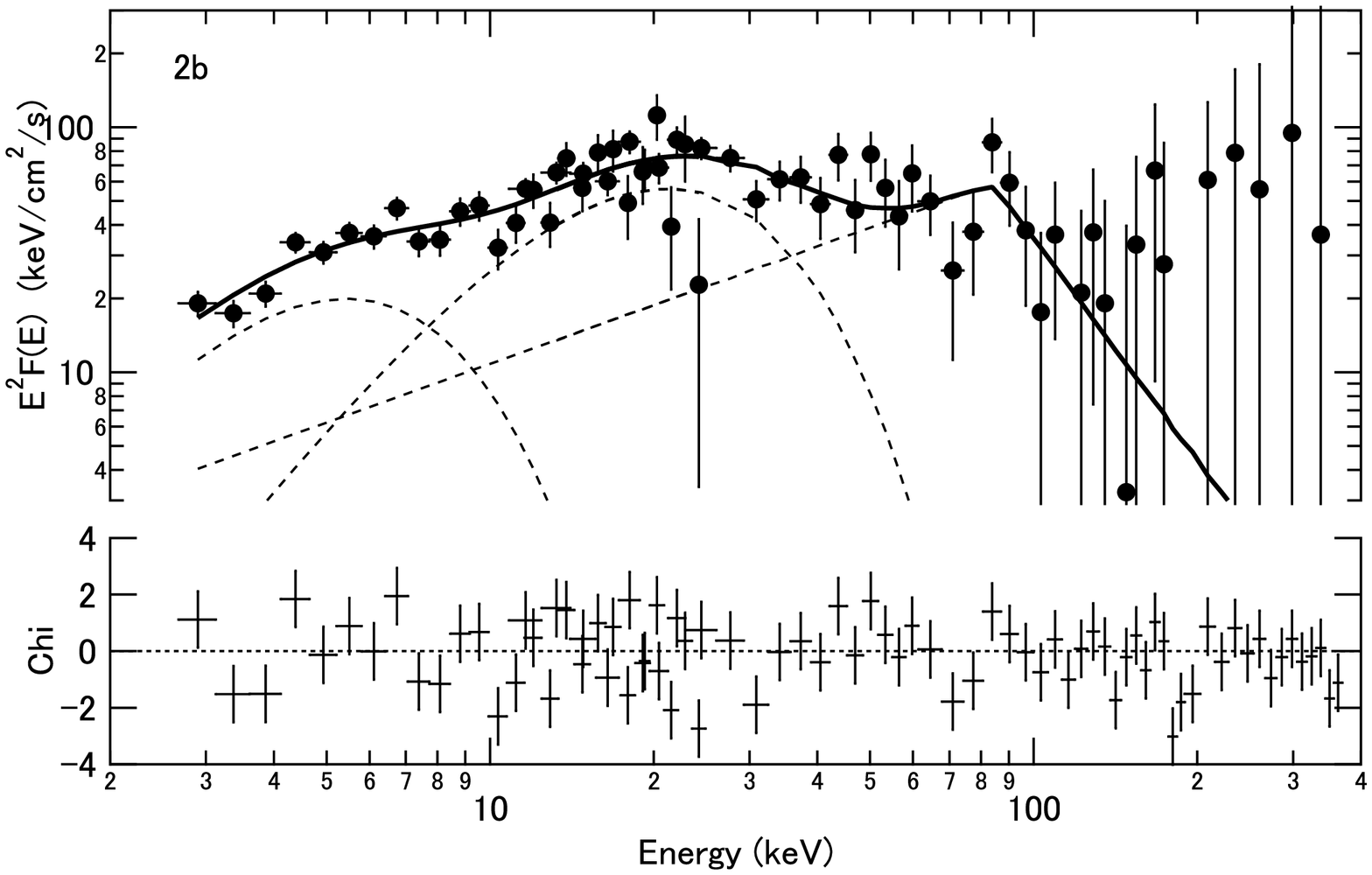} \\
     \FigureFile(0.45\textwidth,0.2\textwidth){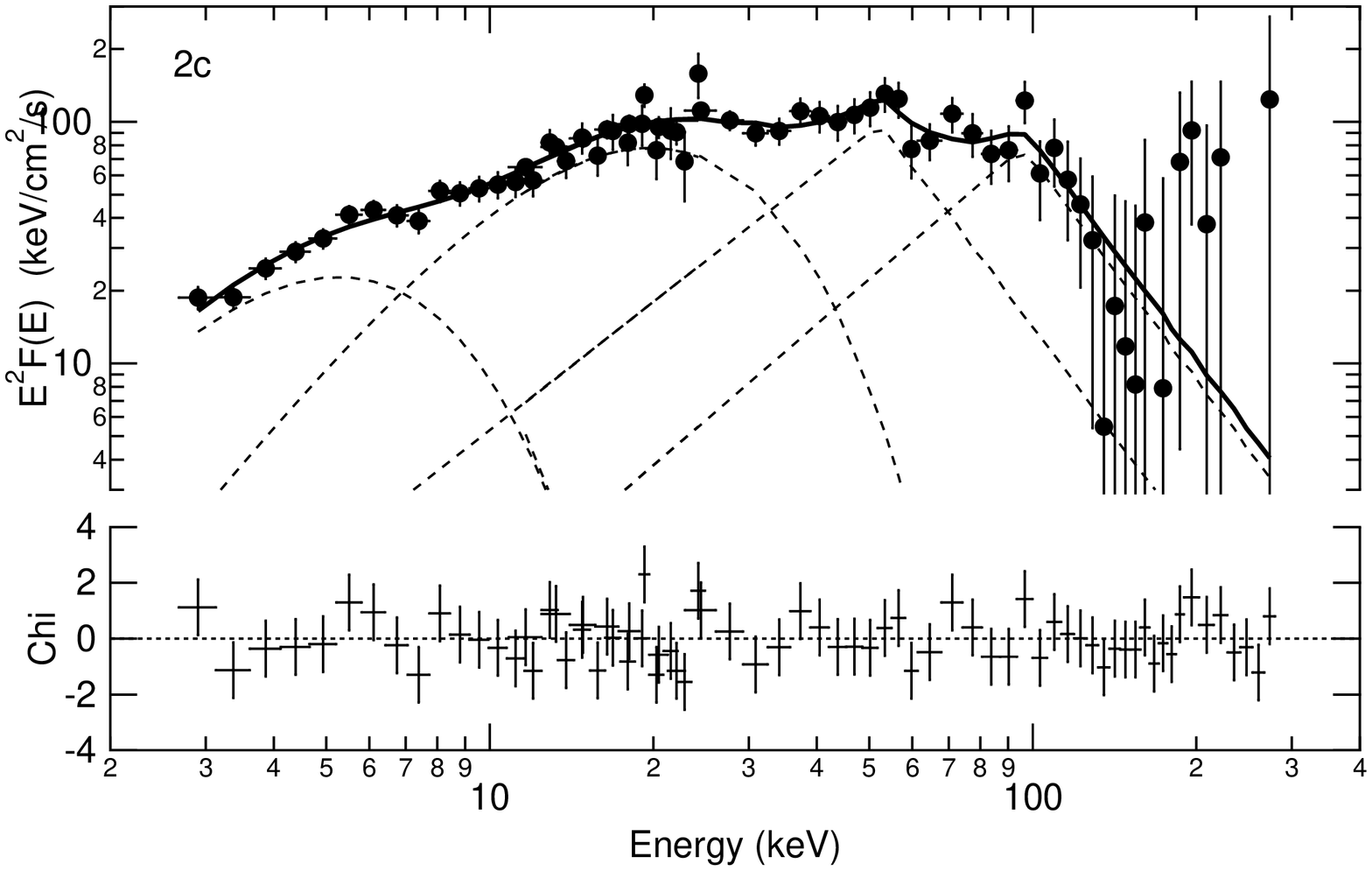} &
     \FigureFile(0.45\textwidth,0.2\textwidth){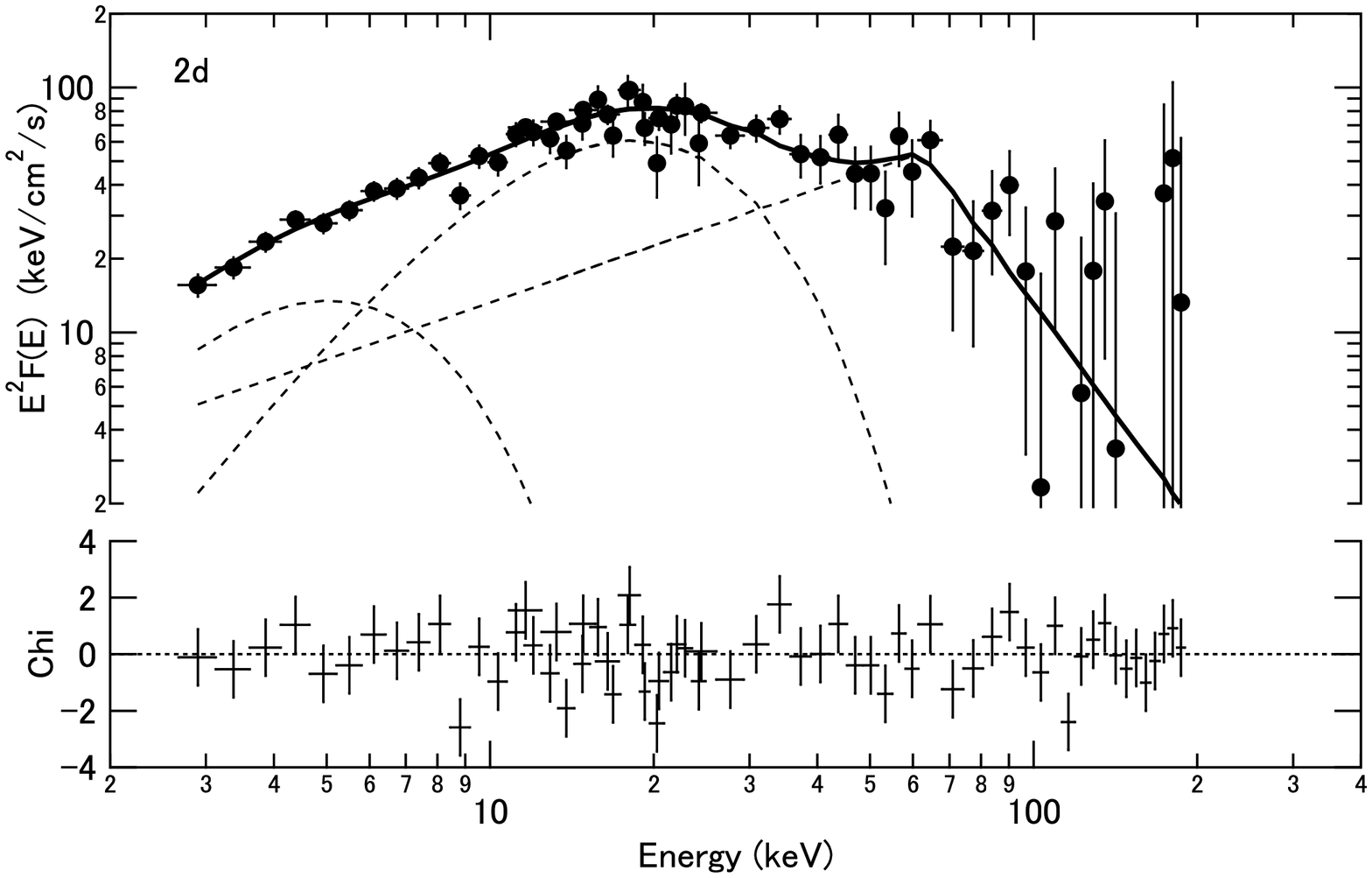} \\
   \end{tabular}
\end{center}
\caption{Time resolved unfolded spectra for intervals~1 and 2.
The residual between the observation and the model is also 
shown at the bottom panel of each figure.
The spectrum is expressed in $\nu f_{\nu}$.
The most preferable model spectra are plotted as a solid line 
(total) and dashed lines (basic function).}
\label{Fig:spect_each1}
\end{figure*}

\begin{figure*}
\begin{center}
   \begin{tabular}{cc}
     \FigureFile(0.45\textwidth,0.2\textwidth){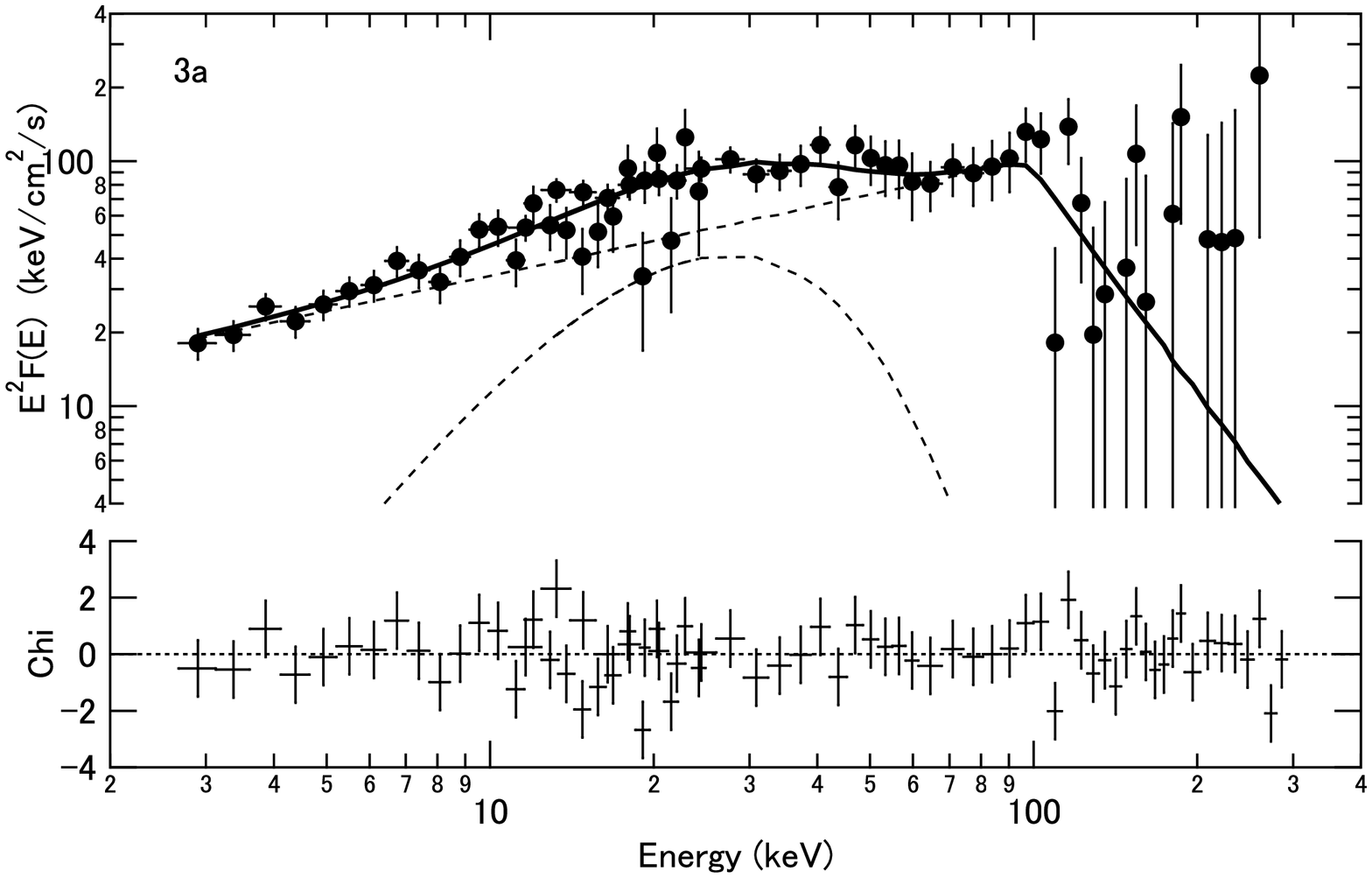} &
     \FigureFile(0.45\textwidth,0.2\textwidth){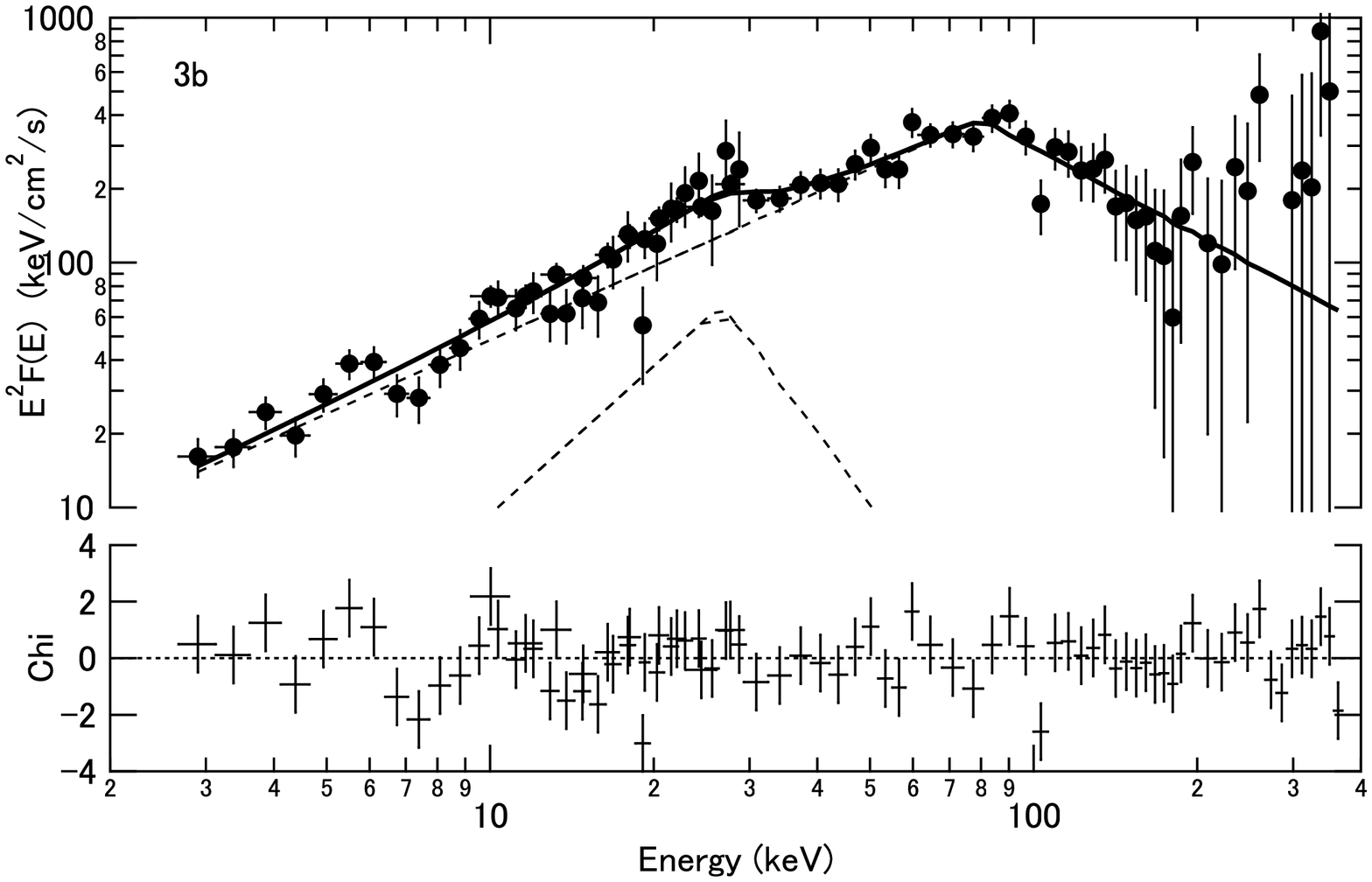} \\
     \FigureFile(0.45\textwidth,0.2\textwidth){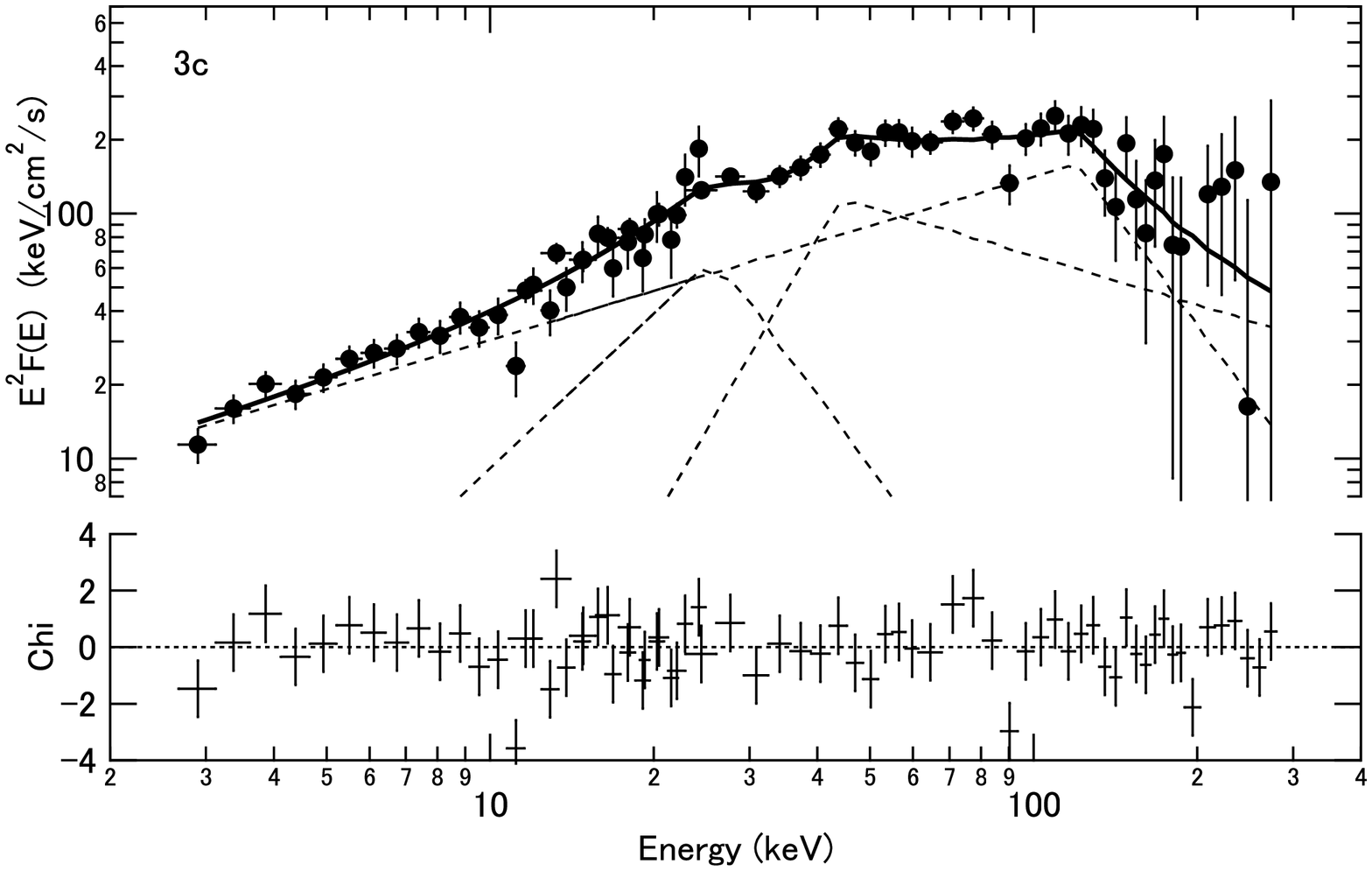} &
     \FigureFile(0.45\textwidth,0.2\textwidth){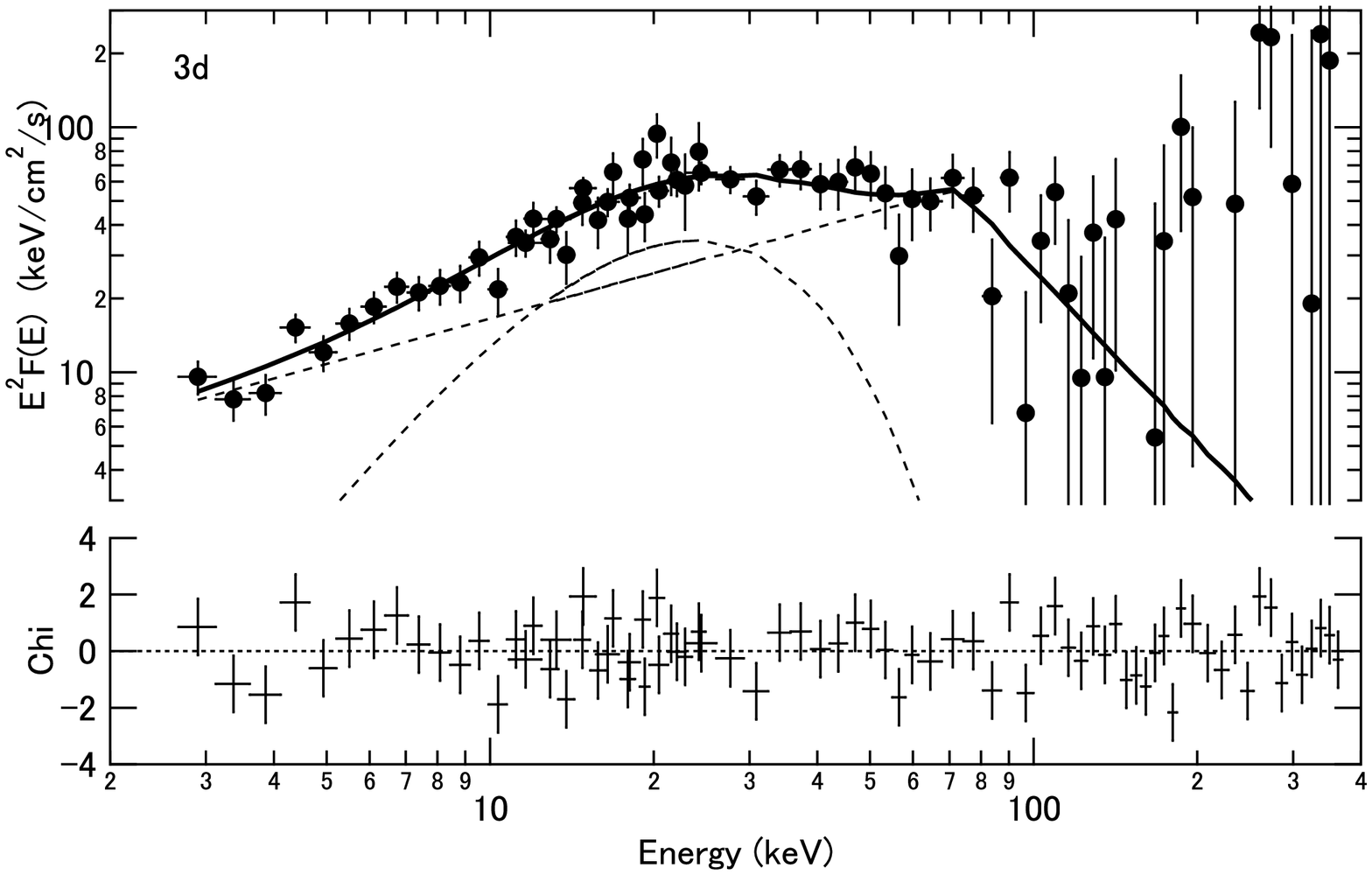} \\
     \FigureFile(0.45\textwidth,0.2\textwidth){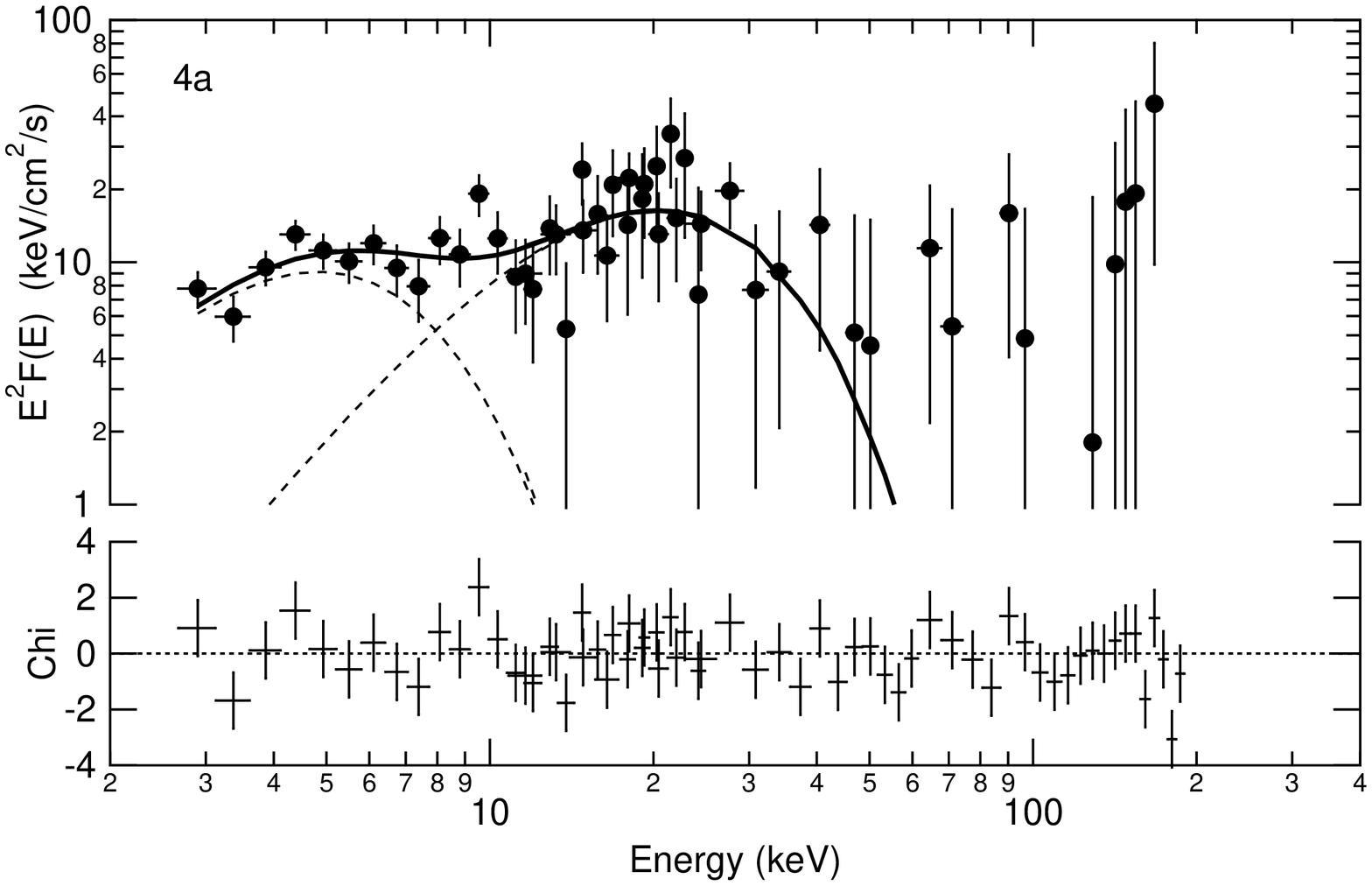} &
     \FigureFile(0.45\textwidth,0.2\textwidth){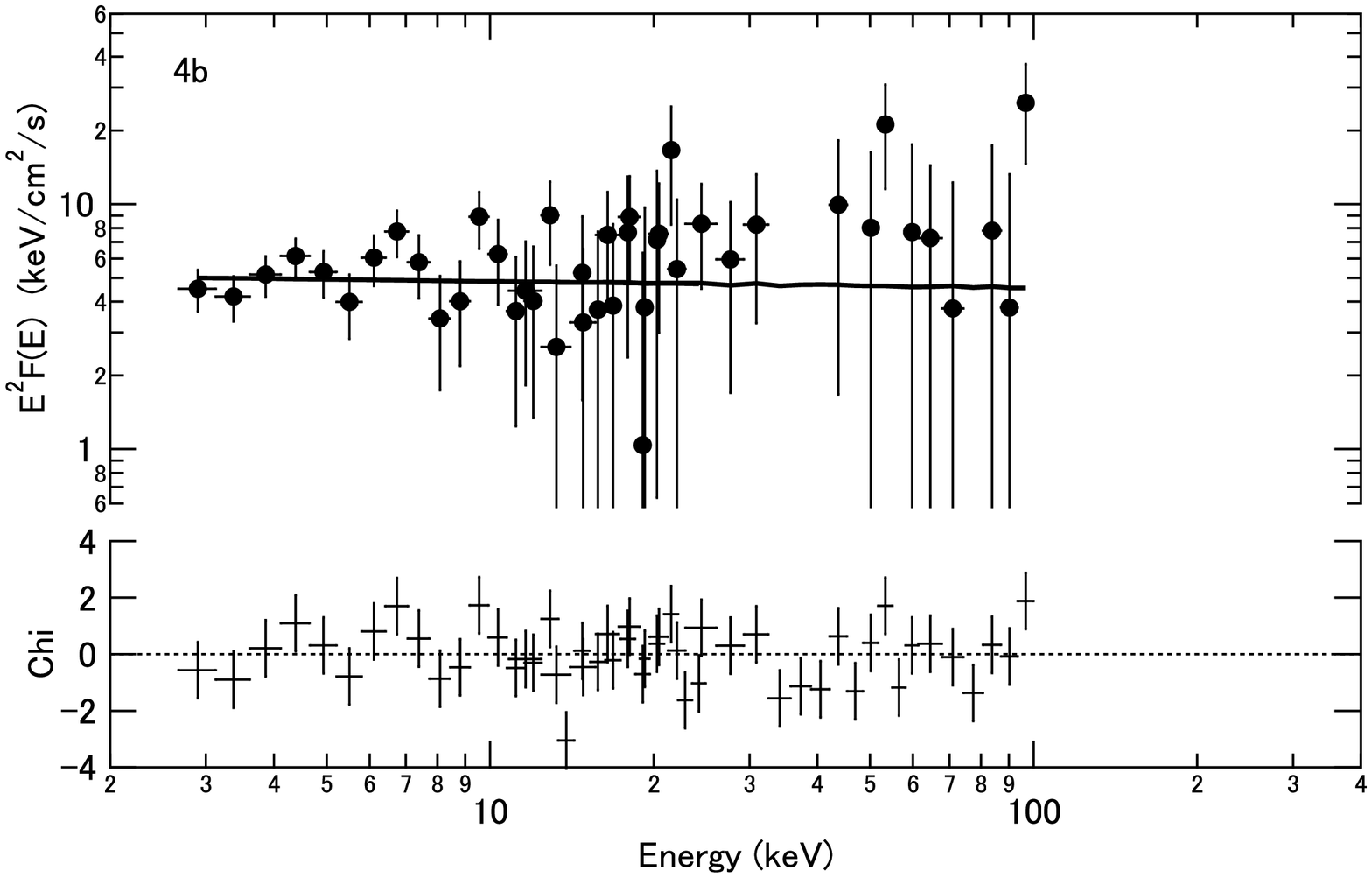} \\
   \end{tabular}
\end{center}
\caption{Time resolved unfolded spectra for interval~3 and 4.}
\label{Fig:spect_each2}
\end{figure*}

The fitting results for various combinations of basic
functions are summarized in table~\ref{tbl:fit_each}.
The fitting parameters for the lowest AIC model are shown in
table~\ref{tbl:fit_params}.
The model spectra giving the lowest AIC at each interval are
shown in figures~\ref{Fig:spect_each1} and \ref{Fig:spect_each2}.
The expected number of components constituting the total
spectrum is inferred from the number of local excesses
in the residual plot for the \verb|bknp| model, and
also from the light curves in the four energy bands.
As an example, the case of interval~2c is shown in 
figure~\ref{fig:spect_2c}.
The spectrum is fitted with a single broken power law function,
and $E_{p}$ is determined as $\sim$20~keV.
Looking at the residual plot shown in the bottom of the figure, 
local excesses around 6~keV and 60~keV are seen.
So the spectrum of interval~2c is expected to be constituted
from three components which have peak energies of 6, 20, and 60~keV.
In the case of interval~2b at least four components are
expected from the light curves.
One is the precursor component seen in interval~1, which
is expected to be present in interval~2 if it is 
extrapolated smoothly.
Two components corresponding to the two peaks seen in 
the 40--80~keV energy band and one component corresponding
to the broad soft emission in the lowest energy band are also 
expected to be present.
So up to four components are examined for interval~2b.

The model selection is carried out by examining the AIC difference,
and the 90\% confidence limit of the AIC difference is calculated by
performing a Monte Carlo simulation.
By this statistical examination, single component models considered
here are rejected for most of the intervals.
The single component model is accepted only for intervals~1a, 
4a, and 4b.
For the other intervals, the single component model considered here
is rejected at 90\% C.L. and the multi-component models are preferred.

For most of the intervals, the null hypothesis probability is larger
than 0.1.
For interval~2b, however, the null hypothesis probability is at most
0.003.
This is probably because unknown systematic errors are present in the
data.

\section{Discussion}

%
%

The optical afterglow light curve in the R band can be 
fitted by a broken power-law model with a break time 
$t_{b} = 0.16 \pm 0.04$ days (\cite{Stanek2005}).
Taking $t_{b}$ as a jet break time
and assuming a homogeneous density profile around the GRB, 
the jet opening angle $\theta$ is estimated from the following 
equation (\cite{Sari1999}, \cite{Nava2006}):
\begin{equation}
\theta = 0.161 \left( \frac{t_{b}}{1+z} \right)^{3/8}
               \left( \frac{n_{0} \eta_{\gamma}}{E_{\rm iso,52}} \right)^{1/8},
\end{equation}
where $n_{0}$ is the ambient particle density in cm$^{-3}$, 
$\eta_{\gamma}$ the radiation efficiency, and $E_{\rm iso, 52} = E_{\rm iso} / $
(10$^{52}$ erg).
Assuming $n_{0} = 3$ and $\eta_{\gamma} = 0.2$,
we obtain a jet opening angle of 3.4$^{\circ}$.
If the GRB is viewed on-axis, the collimation-corrected total energy 
can be estimated from $E_{\gamma} = ( 1 - \cos{\theta} ) E_{\rm iso}$.
The corrected total energies for the three components are
$2.4^{+0.70}_{-1.4} \times  10^{49}$ erg for  $E_{\rm p,src} = 123^{+28}_{-17}$~keV,
$0.49^{+1.8}_{-0.2} \times 10^{49}$ erg for $E_{\rm p,src} = 44^{+3.4}_{-6.9}$~keV,  and
$1.7^{+2.8}_{-1.1} \times  10^{48}$ erg for $E_{\rm p,src} = 8.4^{+2.2}_{-1.0}$~keV.
These values do not follow the Ghirlanda relation (\cite{Ghirlanda2007})
except for the component with $E_{p} \sim $ 6~keV.
That is, the $E_{\rm p,src}$ expected from the Ghirlanda relation
are 39.4, 13.0 and 6.2~keV for the components with $E_{p} >$ 40~keV, $\sim$ 20~keV,
and $\sim$ 6~keV, respectively.
Taking a 5\% uncertainty in the Ghirlanda relation, the observed
$E_{p}$ for the the components with $E_{p} >$ 40~keV and $\sim$ 20~keV
are incompatible.

We also tested the Liang-Zhang relation (\cite{Liang2005}).
The isotropic energies $E_{\rm iso,52}$ calculated by equation~(5) of 
\cite{Liang2005} are:
2.54, 0.132, 3.28 and 24.1 for components ``total'', A, B, 
and C, respectively.
The isotropic energy derived from the fit to a single broken power law function 
are consistent with the isotropic energy derived from the Liang-Zhang 
relation.
On the other hand, the isotropic energies derived for components B and C are 
incompatible with those obtained from the relation.

\begin{figure*}
  \begin{center}
    \FigureFile(1\textwidth,\textwidth){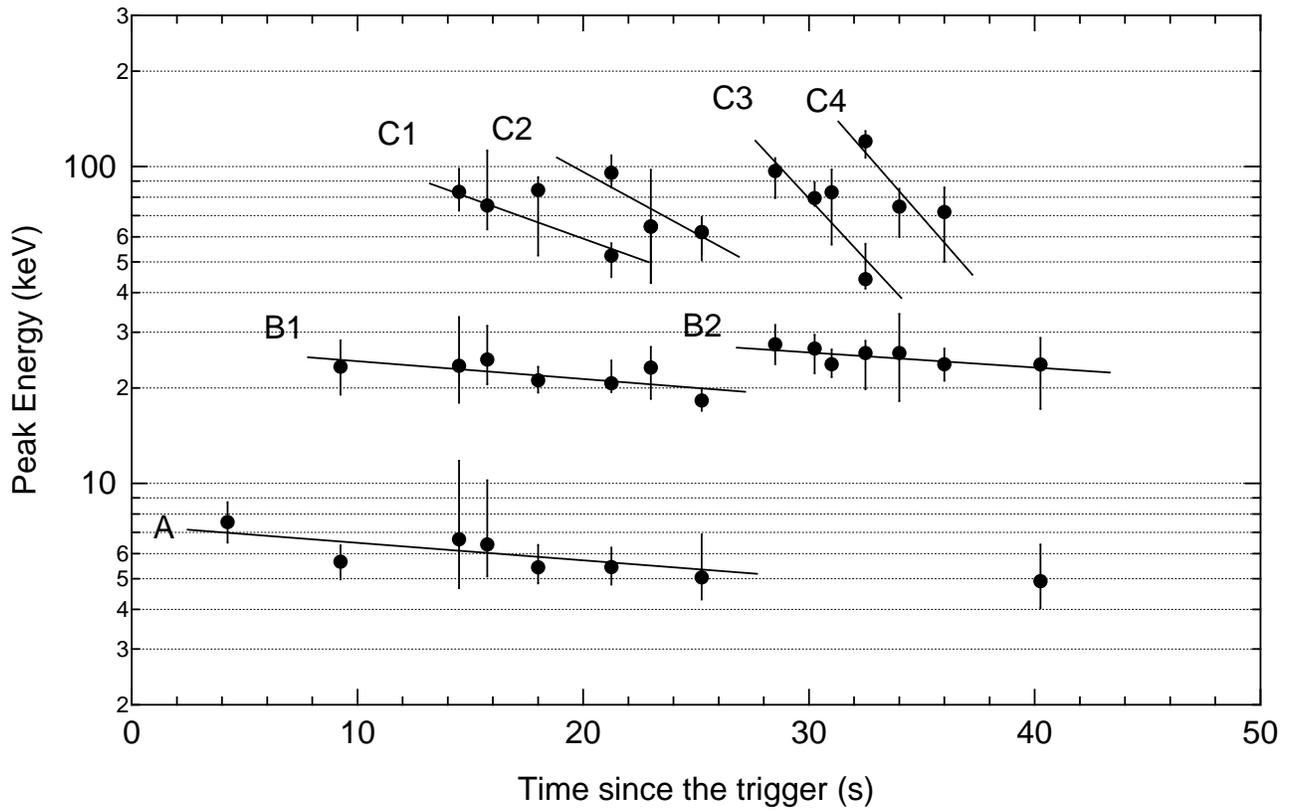}
  \caption{Peak energy calculated for each interval by fitting the data
with multi-component models. 
The points which are inferred to belong to identical components
are interpolated with a line. The vertical error bar corresponds to 
90\% C.L.}
  \label{fig:Epeak}
  \end{center}
\end{figure*}

Looking at the time evolution of $E_{p}$ obtained by the
time resolved spectral analysis shown in figure~\ref{fig:Epeak}, 
we can identify seven components.
Each component is interpolated with a solid line,
and is given an identifier A, B$_{1}$, B$_{2}$, C$_{1}$, C$_{2}$,
C$_{3}$ or C$_{4}$.

The most preferred spectral model for component A in interval~1a 
is the \verb|bbody| model.
The calculated emission radius is $4.35_{-1.1}^{+1.4} \times 10^{6}$~km,
which corresponds to 6 solar radii and is a typical radius for a
blue supergiant.
The AIC difference for the second-most preferred \verb|bknp| model is
3.31 and its 90\% confidence limit is 4.9, so the \verb|bknp| is
also acceptable.
The AIC differences for the power law spectrum with and without 
absorption (\verb|wabs*pl| and \verb|pl|) are larger than 8.9,
and their 90\% confidence limits are less than 0.3, so these models are 
rejected at 90\% C.L.

For interval~1b, the acceptable models are 
\verb|bbody*2|, \verb|bbody+bknp| and \verb|bknp*2|, 
all of which are two-component models.
None of the single component models considered here is
preferable and all are rejected at 90\% C.L.
Thus it is likely that the emission in interval~1b is composed
of two components (A and B$_{1}$).
The spectral type of each component is not uniquely determined from this
result; it is either a black body or a broken power law function.
Assuming that component B$_{1}$ is black body radiation,
the calculated emission radius is about one solar radius.

\begin{figure*}
  \begin{center}
    \FigureFile(0.8\textwidth,0.8\textwidth){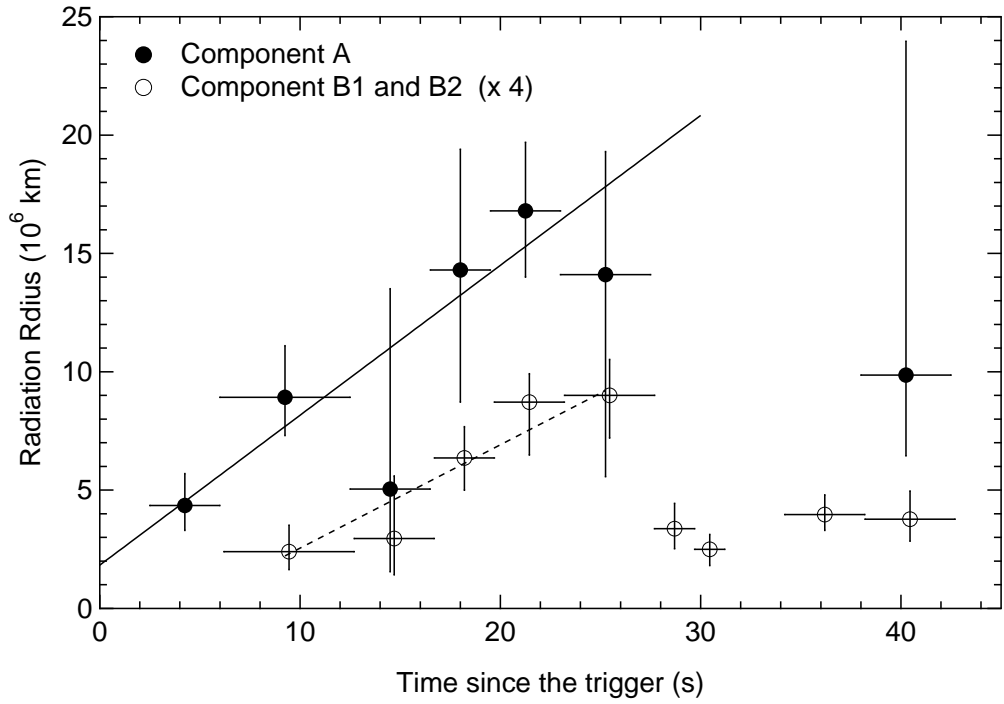}
    \caption{Evolution of the radiation radii of the black body
components. The filled circles represent component A of  
figure~\ref{fig:Epeak}.
The open circles represent components B$_{1}$ and B$_{2}$,
for which the radius is multiplied by four.
The solid and dashed lines represent the linear fit to the data of 
intervals 1 and 2.}
    \label{fig:radius}
  \end{center}
\end{figure*}

In intervals 2a--2d, the soft components A and B$_{1}$
are present in all the acceptable models.
The peak energies of the components are almost constant during intervals
1 and 2, and they decrease slowly, with decay time $72 \pm 42$~sec
for component A and $57 \pm 33$~sec for component B$_{1}$.
Assuming that the components originate from thermal emission, 
we can derive the evolution of the radiation radii, and they are shown in 
figure~\ref{fig:radius} with the filled circles for component A and
with open circles for component $B_{1}$.
The data points for component $B_{1}$ are shifted by a factor of four.
The data points for intervals~1 and 2 are fitted with a linear function,
and we calculate the apparent expansion velocity for component A to be
($6.3 \pm 1.5 ) \times 10^{5}$~km s$^{-1}$, which is twice the speed of light.
This superluminal motion is observed when the emitter is moving 
with relativistic velocity toward the observer.
The relation between the apparent expansion velocity $v$ and
the velocity measured in the source frame $v'$ is given by:
\begin{equation}
   v = \frac{v'}{(1 + z) (1 - \frac{v'}{c})}.
\end{equation}
The expansion velocity in the source frame is $2.35 \times 10^{5}$~km s$^{-1}$,
and the corresponding Lorenz factor is 1.6.
The apparent expansion rate for component B$_{1}$ is found to be
$1.1 \times 10^{5}$~km s$^{-1}$, and the velocity in the source frame is
$1.2 \times 10^{5}$~km s$^{-1}$, which is half the velocity of
component A.
%
%
This result indicates that the soft component originates from the GRB
photosphere expanding with a mildly relativistic speed.
According to the current models of GRB photosphere 
(e.g. \cite{Meszaros2002}; \cite{Rees2005}), however, it is difficult to 
interpret a blackbody with essentially the same temperature but an increasing
radius, unless the temperature is boosted by the growing Lorentz factor
of the photosphere.

\begin{figure*}
  \begin{center}
    \FigureFile(0.8\textwidth,0.8\textwidth){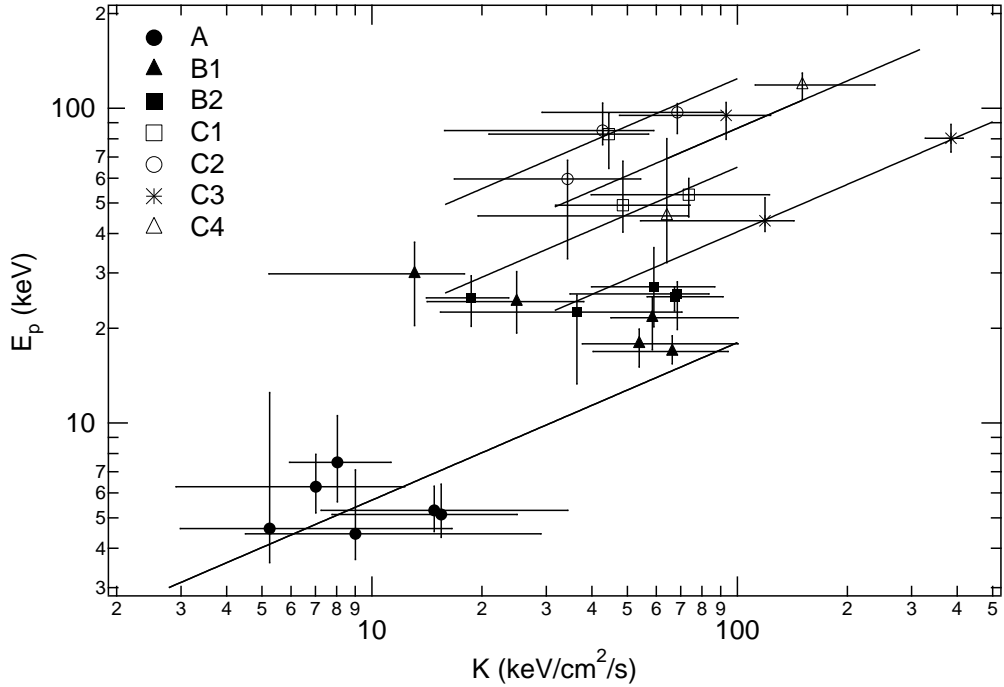}
    \caption{The relation between $E_{p}$ and $K$ of equation~\ref{eq:bknp}
for each component. Solid lines represent the relation $E_{p} \propto K^{0.5}$.
}
    \label{fig:ep_vs_k}
  \end{center}
\end{figure*}

If the component originates in an internal shock according to the
model of \cite{Zhang2002} the following relation should
be satisfied:
\begin{equation}
E_{p} \propto L^{1/2} \Gamma^{-2} 
\label{eq:ep-luminosity}
\end{equation}
where $L$ is the luminosity and $\Gamma$ is the bulk Lorentz factor of 
the shock.
If the spectral shape does not change, the normalization constant $K$ of
equation~\ref{eq:bknp} is proportional to the luminosity.
As the $\alpha$ and $\beta$ are not well constrained in the multi-component 
model due to the correlation of the parameters among the components, the 
luminosity is not well constrained.
We have plotted the $E_{p}$-$K$ relation in figure~\ref{fig:ep_vs_k}.
If $\Gamma$ is constant and the spectral shape does not change during the 
emission, we expect that $E_{p}$ will be proportional to $K^{1/2}$.
No clear correlation is found for component A (filled circle).
For component $B_{1}$ (filled triangle) the expected correlation is 
not found either, and it shows a negative correlation.
%

%
%
%
%
%
%

The higher energy components of the interval~2, C$_{1}$ and C$_{2}$, which 
correspond to the two peaks seen in the 40--80~keV light curve,
are resolved as a broken power law spectrum for which $E_{p}$ is 
around 50--90~keV.
%
%
If we assume that $E_{p}$ decreases exponentially as seen in many
GRBs, we can derive the correspondence among the $E_{p}$ as indicated 
in figure~\ref{fig:Epeak}.
The decay constant of the $E_{p}$ is $\sim$20 sec.

At interval~3, the first precursor component seen
in interval~1a (component A) is not well resolved.
Component B$_{2}$ has a similar $E_{p}$
to that of component B$_{1}$, but its $E_{p}$ is somehow systematically 
higher than the extrapolation of B$_{1}$.
Assuming that B$_{2}$ is thermal emission, its radiation radius
is calculated and shown in figure~\ref{fig:radius}.
The radiation radius is well below the extrapolation of those for B$_{1}$.
The $E_{p}$-$K$ relation of B$_{2}$ is shown in figure~\ref{fig:ep_vs_k},
and it does not follow the relation given by equation~\ref{eq:ep-luminosity}.
The highly variable spectra whose emission peaks vary from 
100~keV to 40~keV are also resolved (C$_{3}$, C$_{4}$), and they correspond to 
the emissions seen in the light curve of the highest energy band.
From figure~\ref{fig:Epeak}, the $E_{p}$ of the components
decrease exponentially with time with a decay constant of
$\sim$5~sec.

The $E_{p}$-$K$ relations for components C$_{1}$, C$_{2}$, C$_{3}$ and 
C$_{4}$ are also shown in figure~\ref{fig:ep_vs_k}.
Although there are few data points for each component, 
the $E_{p}$-$K$ relation is satisfied except for two points.
Both the exceptions are at the time intervals corresponding to the rising
part of the components C$_{1}$ and C$_{3}$.
During the rise, due to the curvature effect, the emission from a part 
of the shock front that is moving toward us dominates.
After that, the emission is averaged over a wider region, so
the emission properties may change between the rising part and the 
following part.

In interval~4a, component B$_{2}$ is likely to remain and
a black body spectrum with $T$ = 1~keV or a broken power law spectrum 
with $E_{p}$ $\sim$ 4~keV is also likely to be present.
In interval 4b, a power law spectrum with photon index 1.9 is
the most preferred model, which is almost the same as the afterglow
spectrum observed by Chandra.

\section{Conclusion}

We have analyzed the time resolved spectra of GRB~041006 and successfully
resolved the components corresponding to the hard spikes and the soft
broad bump observed in the multi-energy band light curves.
The components may be divided into two classes.
One is component A, which has almost constant $E_{p}$ 
around 6~keV, and components B$_{1}$ and B$_{2}$ which have almost 
constant $E_{p}$ around 20~keV.
$E_{p}$ for this class gradually decreases on a timescale, 60--70~s.
The spectral type is well represented by a broken power law function 
or a black body radiation function.
Assuming that the emission of this component is due to black body radiation, 
we derived the emission radii.
At the beginning of the emission they are 4$\times$10$^{6}$~km
for component A and 7$\times$10$^{5}$~km for components B$_{1}$ and B$_{2}$.
The expansion velocity in the source frame is also derived; it is
0.78~c and 0.4~c for components A and B$_{1}$, respectively.
The emission radius of component B$_{2}$ is almost constant.

The $E_{p}$-Luminosity relation is examined for these components and compared
with the prediction of the internal shock model.
We used a normalization constant $K$ in equation~\ref{eq:bknp} instead of 
deriving the luminosity.
According to the internal shock model of ~\cite{Zhang2002}, $E_{p}$
is proportional to $L^{1/2}$ if the bulk Lorentz factor of the shock is 
constant during the emission.
We could not find such a correlation for components A, B$_{1}$ and B$_{2}$.

The second class comprises the components whose $E_{p}$ is larger than 
the former class and shows a relatively rapid decrease on a timescale 
of 5--20 sec.
The spectra are well represented by a broken power law function, and
the $E_{p}$-$K$ relation almost follows the relation expected for an
internal shock origin,  so this could explain their origin.
%
%

We could not reach a conclusion about the origin of the soft component
observed for GRB~041006.
However, the difference in its time variability with respect to 
the higher energy component suggests that it originates
from different emission sites, such as acceleration by a wider jet,
emission from a supernova shock breakout, or emission from the photosphere 
of the fireball.

\section*{Acknowledgements}

We would like to thank the HETE-2 team members for their support. 
The HETE-2 mission is supported in the US by NASA contract NASW-4690; 
in Japan in part by Grant-in-Aid 14079102 from the Ministry of Education, 
Culture, Sports, Science, and Technology; 
and in France by CNES contract 793-01-8479.
YS is grateful for support under the JSPS Core-to-Core Program,
Grant-in-aid for Information Science (15017289 and 18049074)
and Young Scientists (B) (17700085) carried out by the Ministry of
Education, Culture, Sports, Science and Technology of Japan.
KH is grateful for support under MIT Contract SC-A-293291,


\begin{table}
  \begin{center}
  \caption{Temporal properties, $T_{50}$ and $T_{90}$, of GRB~041006.}

  \begin{tabular}{r@{ -- }l c@{$\pm$}c c@{$\pm$}c} \hline
\multicolumn{2}{c}{energy range} & 
\multicolumn{2}{c}{T50\footnotemark[$*$]} & 
\multicolumn{2}{c}{T90\footnotemark[$*$]} \\
\multicolumn{2}{c}{(keV)} & 
\multicolumn{2}{c}{(s)} & 
\multicolumn{2}{c}{(s)} \\ \hline 
2  & 10  & 13.9 & 0.08 & 38.2 & 0.40 \\
10 & 25  & 11.9 & 0.16 & 27.3 & 1.44 \\
40 & 80  & 10.2 & 0.09 & 19.6 & 0.10 \\
80 & 400 &  3.7 & 0.25 & 17.4 & 0.25 \\ \hline
\multicolumn{6}{@{}l@{}}{
   \hbox to 0pt{\parbox{85mm}{\footnotesize
   \par\noindent
   \footnotemark[$*$] The quoted errors correspond to one sigma.
   }\hss}}
  \end{tabular}
  \label{tbl:T50and90}
  \end{center}
\end{table}

\begin{table}
\begin{center}
\caption{Time intervals used for time resolved spectral analysis. }
\begin{tabular}{cr@{ -- }lc}\hline
time interval & start\footnotemark[$*$] & end\footnotemark[$*$] \\ 
              & \multicolumn{2}{c}{(s)} \\ \hline
1a &  2.5 &  6.0 \\
1b &  6.0 & 12.5 \\
2a & 12.5 & 16.5 \\
2b & 16.5 & 19.5 \\
2c & 19.5 & 23.0 \\
2d & 23.0 & 27.5 \\
3a & 27.5 & 29.5 \\
3b & 29.5 & 31.0 \\
3c & 31.0 & 34.0 \\
3d & 34.0 & 38.0 \\
4a & 38.0 & 42.5 \\
4b & 42.5 & 60.0 \\ \hline
2a' & 15.0 & 16.5 \\
2c' & 22.0 & 24.0 \\
3b' & 30.0 & 32.0 \\
3c' & 33.0 & 35.0 \\ \hline
\multicolumn{3}{@{}l@{}}{
   \hbox to 0pt{\parbox{85mm}{\footnotesize
   \par\noindent
   \footnotemark[$*$] 
The offset time is the trigger time 2004-10-06 12:18:08.63933.
   }\hss}}
\end{tabular}
\label{tbl:timeRegion}
\end{center}
\end{table}

%
%
\begin{table*}
\begin{center}
\caption{Results of the spectral fit to the time averaged spectrum.}
\begin{tabular}{lccrrrrl}
\hline
model      & $n$\footnotemark[$*$]
           & $k$\footnotemark[$\dagger$]
           & $\chi^{2}$\footnotemark[$\ddagger$]
           & $p$\footnotemark[$\S$]
           & $AIC$\footnotemark[$\|$]
           & $\Delta_{X}$\footnotemark[$\#$]
           & parameters\footnotemark[$**$]\\ 
\hline
bbody*2+bknp & 83 & 8 & 74.35 & 0.499 &  6.87 & --        & $T$=1.4,5.5,$E_{p}$=74   \\
bknp*3       & 83 &12 & 68.84 & 0.551 &  8.47 & 1.6(4.7) & $E_{p}$=5,25,72   \\
bbody+bknp*2 & 83 &10 & 73.75 & 0.453 & 10.19 & 3.32(4.1) & $T$=1.6,$E_{p}$=23,73  \\
bknp*2       & 83 & 8 & 77.80 & 0.390 & 10.63 & 3.76($<$0) & $E_{p}$=5,24   \\
band         & 83 & 4 & 96.55 & 0.087 & 20.55 &13.68($<$0) & $E_{p}$=38 \\
bknp         & 83 & 4 &111.19 & 0.010 & 32.27 &25.40($<$0) & $E_{p}$=22 \\
\hline
\multicolumn{3}{@{}l@{}}{
   \hbox to 0pt{\parbox{160mm}{\footnotesize
\par\noindent
\footnotemark[$*$] Numbers of data points used for the fit.
\par\noindent
\footnotemark[$\dagger$] Numbers of model parameters.
\par\noindent
\footnotemark[$\ddagger$] Chi-square of the fit
\par\noindent
\footnotemark[$\S$] Null hypothesis probability.
\par\noindent
\footnotemark[$\|$] Akaike information criterion.
\par\noindent
\footnotemark[$\#$] AIC difference between the corresponding model 
and the lowest AIC model. The numbers in parentheses represent the 
90\% confidence limits of the AIC difference.
\par\noindent
\footnotemark[$**$] $T$ 
is the black body temperature in keV and $E_{p}$ is the break energy of
the \texttt{brknp} model in keV. 
   }\hss}}
\end{tabular} 
\label{tbl:fit_all}
\end{center}
\end{table*}

\begin{table*}
\begin{center}
\caption{Fitting parameters for the time averaged spectrum.
 }
\begin{tabular}{ccllll} \hline
model & component & parameters\footnotemark[$*$]\\ \hline
bbody*2+bknp & 1 & $T = 1.40_{-0.16}^{+0.22}$ &
                   $K_{\rm bbody}  = 0.16 \pm 0.04 $ \\
             & 2 & $T = 5.53_{-0.67}^{+0.77}$ &
                   $K_{\rm bbody} = 0.44 \pm 0.10$   \\
             & 3 & $E_{p} = 73.5_{-15.6}^{+7.6}$       
                 & $\alpha = 1.33_{-0.14}^{+0.09}$
                 & $\beta = 2.96_{-0.60}^{+1.19}$
                 & $K_{\rm bknp} = 37.8_{-6.1}^{+6.2}$
 \\
\hline
bknp*3
             & 1 & $E_{p} = 71.9_{-9.6}^{+16}$       
                 & $\alpha = 1.3_{-3.3}^{+0.2}$
                 & $\beta = 2.9_{-0.4}^{+1.2}$
                 & $K_{\rm bknp} = 43.4_{-27}^{+3.5}$ \\
             & 2 & $E_{p} = 25.4_{-4.0}^{+2.0}$       
                 & $\alpha = 1.2_{-0.9}^{+0.3}$
                 & $\beta = 5.00_{-2.5}^{+0.0}$
                 & $K_{\rm bknp} = 19.8_{-3.3}^{+24}$ \\
             & 3 & $E_{p} = 4.9_{-0.6}^{+1.3}$       
                 & $\alpha = -2.00_{-0.0}^{+3.0}$
                 & $\beta = 2.9_{-0.4}^{+2.1}$
                 & $K_{\rm bknp} = 3.69_{-1.0}^{+5.2}$ \\
\hline
\multicolumn{3}{@{}l@{}}{
   \hbox to 0pt{\parbox{160mm}{\footnotesize
\par\noindent
\footnotemark[$*$] $T$ and $K_{\rm bbody} = R_{\rm km}^{2}/D_{10}^{2}$ are 
the temperature in units of keV and normalization constant for the black-body 
radiation model, respectively.
$R_{\rm km}$ is the source radius in units of km.
$D_{10}$ is the distance to the source in units of 10 kpc.
$E_{p}$, $\alpha$, $\beta$, $K_{\rm bknp}$ are the break energy in units of keV, 
low energy photon index, high energy photon index, and normalization constant 
defined in equation~\ref{eq:bknp}.
$K_{\rm bknp}$ is in units of keV cm$^{-2}$ s$^{-1}$.
   }\hss}}
\end{tabular}

\label{tbl:fit_params_total}

\end{center}
\end{table*}

\begin{longtable}{clccrrrrl}
\hline
interval & 
model    & 
$n$\footnotemark[$*$] & 
$k$\footnotemark[$\dagger$] & 
$\chi^{2}$\footnotemark[$\ddagger$] &
$p$\footnotemark[$\S$]& 
$AIC$\footnotemark[$\|$]&
$\Delta_{X}$\footnotemark[$\#$]& 
parameters\footnotemark[$**$] \\ \hline
\endhead
\caption{Results of spectral model fitting to the time resolved
spectra.
} \\
\label{tbl:fit_each}
\endfoot
1a
         & bbody      & 52 & 2 & 41.38 & 0.802 & -7.87 & --        & $T$=2   \\
         & bknp       & 52 & 4 & 40.75 & 0.762 & -4.68 & 3.19(3.9) & $E_{p}$=7.3  \\
         & wabs*pl    & 52 & 3 & 47.26 & 0.544 &  1.03 & 8.90(1.1) & $\alpha$=3.0,$n_{\rm H}$=16   \\
         & pl         & 52 & 2 & 56.57 & 0.243 &  8.38 &16.25($<$0) & $\alpha$=2.1   \\
\hline
%
%
%
1b
         & bbody*2    & 52 & 4 & 36.27 & 0.893 &-10.73 & --       & $T$=1.4,5.9 \\
         & bbody+bknp & 52 & 6 & 35.92 & 0.857 & -7.24 & 3.49(4.2)& $T$=1.5,$E_{p}$=30 \\
         & bknp*2     & 52 & 8 & 35.60 & 0.813 & -3.70 & 7.03(7.4)& $E_{p}$=6,30 \\
         & bknp       & 52 & 4 & 42.92 & 0.681 & -1.98 & 8.75($<$0)& $E_{p}$=6 \\

         & bbody+pl   & 52 & 4 & 49.93 & 0.396 &  5.89 &16.62($<$0)& $T$=2.1,$\alpha$=1.9 \\
         & pl         & 52 & 2 & 63.52 & 0.095 & 14.41 &25.14($<$0)& p=1.9 \\
\hline

%
%
2a
         & bbody*2+bknp & 80 & 8 & 59.34 & 0.857 & -7.90 & --       & $T$=1.7,5.9,$E_{p}$=84 \\
         & bknp*2       & 80 & 8 & 61.24 & 0.813 & -5.38 & 2.52(4.1)& $E_{p}$=24,83 \\
         & bbody+bknp*2 & 80 & 10& 58.43 & 0.837 & -5.14 & 2.76(4.2)& $T$=2.6,$E_{p}$=23,83 \\
         & bknp*3       & 80 & 12& 57.68 & 0.810 & -2.17 & 5.73(9.4)& $E_{p}$=5,24,83 \\
         & bknp         & 80 & 4 & 70.48 & 0.657 & -2.13 & 5.77(0.5)& $E_{p}$=25 \\
\hline
2b
         & bbody*2+bknp & 80 & 8 &104.91 & 0.007 & 37.69 & --        & $T$=1.4,5.4,$E_{p}$=84 \\
       & bbody*2+bknp*2 & 80 & 12& 99.33 & 0.008 & 41.31 & 3.77(6.2) & $T$=1.4,5.5,$E_{p}$=50,85 \\
         & bbody+bknp   & 80 & 6 &116.18 & 0.001 & 41.85 & 3.99(2.0) & $T$=1.5,$E_{p}$=21 \\
         & bknp         & 80 & 4 &122.30 & 0.001 & 41.96 & 4.10(1.7) & $E_{p}$=23 \\
         & bknp*2       & 80 & 8 &111.59 & 0.002 & 42.63 & 4.77(4.1) & $E_{p}$=23,85 \\
         & bknp*3       & 80 & 12&101.08 & 0.006 & 42.71 & 4.78(8.2) & $E_{p}$=5,22,85 \\
         & bbody+bknp*2 & 80 & 10&106.05 & 0.004 & 42.55 & 5.22(5.5) & $T$=1.5,$E_{p}$=22,85 \\
\hline
2c
       & bbody*2+bknp*2 & 73 & 12& 49.53 & 0.853 & -4.32 & --         & $T$=1.3,5.0,$E_{p}$=52,98 \\
         & bbody*2+bknp & 73 & 8 & 56.66 & 0.760 & -2.50 & 1.67($<$0) & $T$=1.3,5.0,$E_{p}$=53 \\
         & bbody+bknp*2 & 73 & 10& 56.61 & 0.702 &  1.44 & 5.76(0.2) & $T$=1.5,$E_{p}$=18,54 \\
         & bknp*3       & 73 & 12& 53.58 & 0.739 &  1.42 & 5.74(0.2) & $E_{p}$=5.5,18,74 \\
         & bknp*2       & 73 & 8 & 62.24 & 0.574 &  4.36 & 8.68(0.06) & $E_{p}$=19,54 \\
         & bbody+bknp   & 73 & 6 & 66.70 & 0.488 &  5.41 & 9.73($<$0)  & $T$=4.7,$E_{p}$=55 \\
         & bknp         & 73 & 4 & 87.99 & 0.006 & 21.63 & 25.72($<$0) & $E_{p}$=23 \\
\hline
2d
         & bbody*2+bknp & 66 & 8 & 64.70 & 0.254 & 14.69 & ----      & $T$=1.2,4.6,$E_{p}$=62 \\
         & bbody+bknp   & 66 & 6 & 72.12 & 0.136 & 17.85 & 3.16(0.9) & $T$=4.5,$E_{p}$=62 \\
         & bknp*2       & 66 & 8 & 70.33 & 0.129 & 20.19 & 5.50(1.2) & $E_{p}$=18,59 \\
         & bknp         & 66 & 4 & 80.21 & 0.060 & 20.87 & 6.18(0.1) & $E_{p}$=18 \\
         & bbody+bknp*2 & 66 & 10& 67.50 & 0.140 & 21.48 & 6.79(5.5) & $T$=1.6,$E_{p}$=17,60 \\
         & bknp*3       & 66 & 12& 66.84 & 0.113 & 24.83 & 10.14(4.4)& $E_{p}$=4,17,60 \\
\hline
3a
         & bbody+bknp   & 74 & 6 & 63.37 & 0.636 & 0.53 & --        & $T$=6.8,$E_{p}$=96 \\
         & bknp*2       & 74 & 8 & 63.72 & 0.557 & 4.93 & 4.40(4.9) & $E_{p}$=27,95 \\
         & bbody+bknp*2 & 74 &10 & 61.83 & 0.554 & 6.71 & 6.18(6.8) & $T$=6.0,$E_{p}$=50,92 \\
         & bknp         & 74 & 4 & 75.48 & 0.306 & 9.46 & 8.93(3.4) & $E_{p}$=36 \\
         & bknp*3       & 74 & 12& 62.21 & 0.469 &11.15 &10.62(11.8)& $E_{p}$=26,45,96 \\
\hline
3b
         & bknp*2       & 84 & 8 & 80.20 & 0.349 & 12.11 & --          & $E_{p}$=25,82 \\
         & bknp*2+pl    & 84 & 10& 79.57 & 0.308 & 15.45 & 3.34(3.9)   & $E_{p}$=26,84,$\alpha$=1.3 \\
         & bbody+bknp+pl& 84 & 8 & 83.64 & 0.257 & 15.64 & 3.53(3.0)   & $T$=8,$E_{p}$=84,$\alpha$=1.6 \\
         & bknp*4       & 84 & 16& 69.19 & 0.437 & 15.69 & 3.58(8.6)   & $E_{p}$=6,10,21,84 \\
         & bbody+bknp*2 & 84 & 10& 80.17 & 0.292 & 16.08 & 3.97(4.0)   & $T$=0.9,$E_{p}$=26,80\\
         & bbody+bknp   & 84 & 6 & 85.91 & 0.413 & 17.89 & 5.78($<$0)& $T$=8,$E_{p}$=83 \\
         & bknp*3       & 84 & 12& 79.88 & 0.245 & 19.78 & 7.67(7.2)   & $E_{p}$=5,26,80 \\
         & bknp         & 84 & 4 &107.35 & 0.022 & 28.60 &16.49($<$0)& $E_{p}$=67 \\
\hline
3c
         & bknp*3       & 73 & 12& 70.36 & 0.193 & 21.32 & --        & $E_{p}$=26,44,120 \\
         & bbody+bknp*3 & 73 & 14& 67.43 & 0.211 & 22.20 & 0.88(4.5) & $T$=1.2,$E_{p}$=26,44,118 \\
         & bknp*2       & 73 & 8 & 80.75 & 0.090 & 23.37 & 2.05(2.3) & $E_{p}$=44,130 \\
         & bbody+bknp*2 & 73 & 10& 78.07 & 0.096 & 24.90 & 3.58(1.3) & $T$=1.1,$E_{p}$=44,117 \\
         & bknp*4       & 73 & 16& 67.91 & 0.153 & 26.72 & 5.40(7.4) & $E_{p}$=6,26,44,119 \\
         & bknp         & 73 & 4 & 98.92 & 0.011 & 30.18 &8.86($<$0)& $E_{p}$=56 \\
\hline
3d
         & bbody+bknp   & 80 & 6 & 76.28 & 0.405 &  8.19 & --        & $T$=6.1,$E_{p}$=72 \\
         & bknp*2       & 80 & 8 & 77.40 & 0.310 & 13.36 & 5.17(5.8) & $E_{p}$=21,47 \\
         & bknp         & 80 & 4 & 86.42 & 0.194 & 14.18 & 5.99($<$0) & $E_{p}$=24 \\
         & bknp*3       & 80 & 12& 74.91 & 0.264 & 18.74 &10.55(13.6)& $E_{p}$=23,43,75 \\
\hline
4a
         & bbody*2      & 66 & 4 & 59.23 & 0.576 &  0.86 &           & $T$=1.2,5.2 \\
         & bbody+bknp   & 66 & 6 & 59.14 & 0.505 &  4.76 & 3.90(7.1) & $T$=1.2,$E_{p}$=24 \\
         & bknp         & 66 & 4 & 63.09 & 0.438 &  5.02 & 4.16(2.8) & $E_{p}$=26 \\
         & bknp*2       & 66 & 8 & 57.36 & 0.496 &  6.74 & 5.88(7.4) & $E_{p}$=4,25 \\
         & bbody+pl     & 66 & 4 & 73.06 & 0.159 & 14.71 &13.85(1.4) & $T$=4.7,$\alpha$=2.3 \\
         & pl           & 66 & 2 &100.05 & 0.003 & 31.46 &30.60($<$0) & $\alpha$=2.0 \\
\hline
4b
         & pl           & 52 & 2 & 47.31 & 0.582 & -0.92 &  --       & $\alpha$=1.9 \\
         & bbody+pl     & 52 & 4 & 44.82 & 0.604 &  0.27 & 1.19(3.1) & $T$=1.5,$\alpha$=1.8 \\
         & bknp         & 52 & 4 & 45.13 & 0.591 &  0.63 & 1.55(3.6) & $E_{p}$=4 \\
         & bbody        & 52 & 2 & 69.71 & 0.034 & 19.24 &20.16($<$0) & $T$=1.7 \\
\hline
\multicolumn{3}{@{}l@{}}{
   \hbox to 0pt{\parbox{160mm}{\footnotesize
\par\noindent
\footnotemark[$*$] Numbers of data points used for the fit.
\par\noindent
\footnotemark[$\dagger$] Numbers of model parameters.
\par\noindent
\footnotemark[$\ddagger$] Chi-square of the fit
\par\noindent
\footnotemark[$\S$] Null hypothesis probability.
\par\noindent
\footnotemark[$\|$] Akaike information criterion.
\par\noindent
\footnotemark[$\#$] AIC difference between the corresponding model 
and the lowest AIC model. The numbers in parentheses represent the 
90\% confidence limits of the AIC difference.
\par\noindent
\footnotemark[$**$] $T$ is the black body temperature
in units of keV, $E_{p}$ is the break energy of the \texttt{brknp} model
in units of keV, $\alpha$ is the power law photon index of the \texttt{pl} 
model, and $n_{\rm H}$ is the column density measured in unit $10^{22}$.
   }\hss}}
\end{longtable}

\begin{table*}
\begin{center}
\caption{Fitting parameters for the most preferred models, that is, the model
that gives the lowest AIC. }
\begin{tabular}{ccllll} \hline
interval & component & parameters\footnotemark[$*$] \\ \hline
1a & 1 & $T = 1.92_{-0.27}^{+0.30}$ &
         $K_{\rm bbody} = 9.94_{-0.42}^{+0.71} \times 10^{1}$ \\
\hline
1b & 1 & $T = 1.44_{-0.17}^{+0.18}$ &
         $K_{\rm bbody} = 4.17_{-1.4}^{+2.2} \times 10^{2}$ \\
   & 2 & $T = 5.94_{-1.08}^{+1.26}$ &
         $K_{\rm bbody} = 1.89_{-0.99}^{+2.1}$ \\
\hline
2a & 1 & $T = 1.60_{-0.21}^{+0.84}$ &
         $K_{\rm bbody} = 2.38_{-2.3}^{+7.1} \times 10^{2}$ \\
   & 2 & $T = 5.75_{-1.2}^{+1.4}$ &
         $K_{\rm bbody} = 3.95_{-3.3}^{+5.9}$ \\
   & 3 & $E_{p} = 83.2_{-10.6}^{+15.2}$ &
         $\alpha = 1.45_{-0.41}^{+0.20}$ &
         $\beta  = 5.00_{-1.8}^{+0.0}$ &
         $K_{\rm bknp} = 48.9_{-11}^{+5.8}$ \\
\hline
2b & 1 & $T = 1.40_{-0.17}^{+0.23}$ &
         $K_{\rm bbody} = 1.02_{-0.63}^{+0.73} \times 10^{3}$ \\
   & 2 & $T = 5.40_{-0.49}^{+0.59}$ &
         $K_{\rm bbody} = 13.0_{-5.7}^{+6.7}$ \\
   & 3 & $E_{p} = 84.3_{-32}^{+8.4}$ &
         $\alpha = 1.26_{-0.83}^{+0.46}$ &
         $\beta = 5.00_{-0.94}^{+0.00}$ &
         $K_{\rm bknp} = 57.8_{-12.2}^{+13.9}$\\
\hline
2c & 1 & $T = 1.34_{-0.077}^{+0.18}$ &
         $K_{\rm bbody} = 1.44_{-0.43}^{+0.56} \times 10^{3}$   \\
   & 2 & $T = 5.01_{-0.46}^{+1.1}$ &
         $K_{\rm bbody} = 25.0_{-13}^{+6.9}$ \\
   & 3 & $E_{p} = 52.3_{-7.6}^{+5.0}$ &
         $\alpha = 0.24_{-2.2}^{+1.0}$ &
         $\beta = 5.00_{-1.9}^{+0.0}$ &
         $K_{\rm bknp} = 97.9_{-40}^{+35}$ \\
   & 4 & $E_{p} = 95.5_{-9.7}^{+13.0}$ &
         $\alpha = 0.06_{-2.1}^{+1.4}$ &
         $\beta = 5.00_{-1.4}^{+0.00}$ &
         $K_{\rm bknp} = 78.4_{-50}^{+19}$ \\
\hline
2d & 1 & $T = 1.28_{-0.19}^{+0.47}$ &
         $K_{\rm bbody} = 1.01_{-0.85}^{+0.95}\times 10^{3}$\\
   & 2 & $T = 4.65_{-0.33}^{+0.42}$ &
         $K_{\rm bbody} = 26.3_{-9.4}^{+9.7}$\\
   & 3 & $E_{p} = 62.1_{-11.5}^{+7.1}$ &
         $\alpha = 1.22_{-1.1}^{+0.3}$ &
         $\beta = 5.00_{-1.4}^{+0.0}$ &
         $K_{\rm bknp} = 54.1_{-10.9}^{+11.5}$ \\
\hline
3a & 1 & $T = 6.8_{-1.1}^{+1.2}$ &
         $K_{\rm bbody} = 3.61_{-1.5}^{+2.9}$ \\
   & 2 & $E_{p} = 95.8_{-15}^{+8.5}$ &
         $\alpha = 1.50_{-0.07}^{+0.07}$ &
         $\beta = 5.00_{-1.5}^{+0.0}$ &
         $K_{\rm bknp} = 107_{-18}^{+17}$\\
\hline
3b & 1 & $E_{p} = 25.3_{-2.6}^{+3.5}$ &
         $\alpha = -0.92_{-1.1}^{+1.5}$ &
         $\beta = 5.00_{-3.2}^{+0.0}$ &
         $K = 68.7_{-11}^{+11}$ \\
   & 2 & $E_{p} = 81.9_{-9.9}^{+7.3}$ &
         $\alpha = 1.05_{-0.10}^{+0.15}$ &
         $\beta = 3.28_{-0.46}^{+0.52}$ &
         $K = 386_{-71}^{+32}$ \\
\hline
3c & 1 & $E_{p} = 25.8_{-4.0}^{+2.4}$ &
         $\alpha = -0.10_{-1.9}^{+0.72}$ &
         $\beta = 5.00_{-2.8}^{+0.0}$ &
         $K = 68.1_{-45}^{+15}$  \\
   & 2 & $E_{p} = 44.0_{-3.6}^{+13}$ &
         $\alpha = -2.00_{-0.00}^{+2.7}$ &
         $\beta = 2.66_{-0.39}^{+2.0}$ &
         $K = 115_{-62}^{+30}$ \\
   & 3 & $E_{p} = 119_{-12}^{+11}$ &
         $\alpha = 1.33_{-0.11}^{+0.05}$ &
         $\beta = 5.00_{-1.40}^{+0.00}$ &
         $K = 159_{-48}^{+95}$ \\
\hline
3d & 1 & $T = 6.05_{-0.69}^{+0.71}$ &
         $K_{\rm bbody} = 5.18_{-1.6}^{+2.4}$ \\

   & 2 & $E_{p} = 71.9_{-30}^{+14}$ &
         $\alpha = 1.39_{-0.10}^{+0.05}$ &
         $\beta = 4.32_{-1.5}^{+0.68}$ &
         $K = 55.7_{-12}^{+12}$ \\
\hline
4a & 1 & $T = 1.23_{-0.16}^{+0.18}$ &
         $K_{\rm bbody} = 8.09_{-3.1}^{+5.6} \times 10^{2}$ \\
   & 2 & $T = 5.16_{-0.71}^{+0.81}$ &
         $K_{\rm bbody} = 4.66_{-2.0}^{+3.5}$ \\
\hline
4b & 1 & $\alpha = 1.93_{-0.14}^{+0.16}$ &
         $K_{\rm pl} = 2.74_{-0.68}^{+0.90}$  \\ \hline
\multicolumn{3}{@{}l@{}}{
   \hbox to 0pt{\parbox{160mm}{\footnotesize
\par\noindent
\footnotemark[$*$] $T$ and $K_{\rm bbody} = R^{2}_{\rm km}/D_{10}^{2}$ are the
temperature in units of keV and normalization constant for the black-body 
radiation model, respectively.
$R_{\rm km}$ is the source radius in units of km.
$D_{10}$ is the distance to the source in units of 10 kpc.
$E_{p}$, $\alpha$, $\beta$, $K_{\rm bknp}$ are the break energy in units of keV, 
low energy photon index, high energy photon index, and normalization constant 
defined in equation~\ref{eq:bknp}. 
The unit of $K_{\rm bknp}$ is  keV cm$^{-2}$ s$^{-1}$.
$K_{\rm pl}$ is the normalization constant for power law spectrum defined as 
photon flux at 1~keV in unit of photons keV$^{-1}$ cm$^{-2}$ s$^{-1}$.
   }\hss}}
\end{tabular}

\label{tbl:fit_params}
\end{center}
\end{table*}


\begin{table}
\begin{center}
\caption{Isotropic energies $E_{\rm iso,52}$ and rest-frame peak energies 
$E_{\rm p,src}$ derived from the average spectrum.}
\begin{tabular}{ccllll} \hline
Component & $E_{\rm p,src}$     & $E_{\rm iso,52}$     \\
          &  (keV)              &                      \\ \hline
Total\footnotemark[$*$]
          & 38.6 $\pm$ 2.9      & 2.54$^{+0.46}_{-0.35}$ \\
A\footnotemark[$\dagger$]
          & 8.4$^{+2.2}_{-1.0}$ & 0.094$^{+0.16}_{-0.08}$ \\
B\footnotemark[$\dagger$]
          & 44$^{+3.4}_{-6.9}$  & 0.28$^{+1.0}_{-0.1}$ \\
C\footnotemark[$\dagger$]
          & 123$^{+28}_{-17}$   & 1.36$^{+0.4}_{-0.8}$   \\ 
C'\footnotemark[$\ddagger$]
          & 126$^{+13}_{-27}$   & 1.32$^{+0.5}_{-0.3}$ \\
\hline

\multicolumn{3}{@{}l@{}}{
   \hbox to 0pt{\parbox{85mm}{\footnotesize
\par\noindent
\footnotemark[$*$] \texttt{bknp} model.
\par\noindent
\footnotemark[$\dagger$] \texttt{bknp*3} model.
\par\noindent
\footnotemark[$\ddagger$] \texttt{bbody*2+bknp} model.
   }\hss}}

\end{tabular}
\label{tbl:ep_eiso}
\end{center}
\end{table}


\clearpage


\clearpage



\clearpage



\end{document}